\documentclass[aip,preprint,amsmath,amssymb]{revtex4-2}
\usepackage{graphicx}

\usepackage{dcolumn}
\usepackage{bm}
\usepackage{hyperref}
\hypersetup{
    colorlinks,
    linkcolor={blue},
    citecolor={blue},
    urlcolor={blue}
}

\begin{document}


\title{Load estimation in unsteady flows from sparse pressure measurements: Application of transition networks to experimental data} 



\author{Giovanni Iacobello}\email{g.iacobello@surrey.ac.uk}
\affiliation{Department of Mechanical and Materials Engineering, Queen's University, Kingston, Ontario, K7L 3N6, Canada}
\affiliation{Department of Mechanical Engineering Sciences, University of Surrey, Guildford, GU2 7XH, UK}
\author{Frieder Kaiser}\email{f.kaiser@queensu.ca}
  \affiliation{Department of Mechanical and Materials Engineering, Queen's University, Kingston, Ontario, K7L 3N6, Canada}
  \author{David E. Rival}
  \affiliation{Department of Mechanical and Materials Engineering, Queen's University, Kingston, Ontario, K7L 3N6, Canada}

\date{\today}

\begin{abstract}

Inspired by biological swimming and flying with distributed sensing, we propose a data-driven approach for load estimation that relies on complex networks. We exploit sparse, real-time pressure inputs, combined with pre-trained transition networks, to estimate aerodynamic loads in unsteady and highly-separated flows. The transition networks contain the aerodynamic states of the system as nodes along with the underlying dynamics as links. A weighted average-based (WAB) strategy is proposed and tested on realistic experimental data on the flow around an accelerating elliptical plate at various angles-of-attack. Aerodynamic loads are then estimated for angles of attack cases not included in the training dataset so as to simulate the estimation process. An optimization process is also included to account for the system's temporal dynamics. Performance and limitations of the WAB approach are discussed, showing that transition networks can represent a versatile and effective data-driven tool for real-time signal estimation using sparse and noisy signals (such as surface pressure) in realistic flows.

\end{abstract}


\maketitle 


\section{Introduction}\label{sec:intro}

\begin{figure*}
	\centering
	\includegraphics[width=0.99\textwidth]{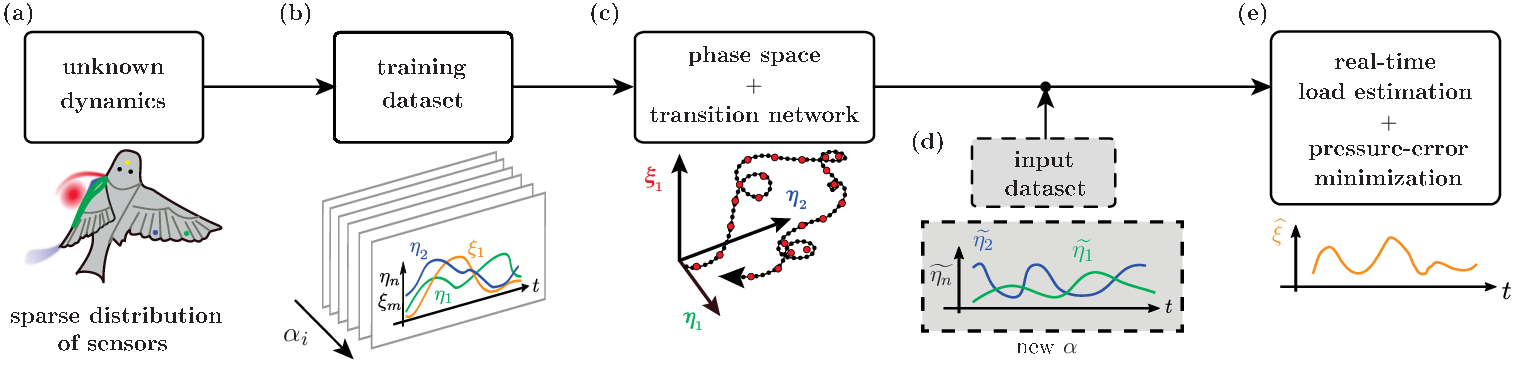}
	\caption{Workflow for the estimation process: (a) Complex (bio-inspired) dynamical system with sparse sensors; (b) Training datased made up of a collection of time series for different configurations, $\alpha_i$ (each collection comprises $N$ time series, $\eta_n(t)$, from sparse sensors and $M$ time series, $\xi_m$, corresponding to the variables to estimate); (c) Phase-space definition of dimension equal to $N+M$, and construction of the transition networks; (d) Collection of a testing dataset of input time series, $\widetilde{\eta}_n$; and (e) Estimation of the $M$ unknown signals, $\xi_m$, using the real-time input in panel (d) and a pre-trained transition networks of panel (c).}	\label{fig:intro2}
\end{figure*}

The instantaneous loads in biological swimming and flying are highly sensitive to environmental perturbations, such as the wakes of other animals, or gusts in the atmosphere, respectively. Despite challenging boundary conditions, animals control the flow over their propulsors (i.e., flippers or wings) with ease and even utilize unsteady flows to their advantage \citep{liao2003fish, portugal2014upwash}. Biological sensory systems monitor the flow in real-time by gathering feedback at multiple locations on the propulsors. By combining the sensor input with their experience, animals instantaneously estimate and control their present aerodynamic state (e.g., the aerodynamic loads) \citep{zbikowski2004, sterbing2011}.
These insights have inspired a series of studies adapting the multi-sensor principle for the control of autonomous swimming and flying vehicles \citep{saini2021leading,provost2018,Burelle2020}. In the absence of simple aerodynamic models for three-dimensional (3D) and highly-separated flows, data-driven methods have been proposed that utilize sparse pressure data to characterize the instantaneous aerodynamic state on an arbitrary body (e.g., a wing) under a variety of conditions (\figurename \ref{fig:intro2}a,b). Examples include attached flow, separated two-dimensional flows, weakly-separated, and highly-separated three-dimensional flows \citep{provost2018, wood2019, hou2019, Burelle2020}.

With the aim to perform (aerodynamic) load estimation, data-driven methods have been shown to be a valid option, although they usually tend to perform well only within a limited range of unsteady boundary conditions~\citep{provost2018, wood2019, hou2019, Burelle2020}. The need for large training datasets, as well as the availability of very sparse pressure distributions, represent a challenge for data-driven methods attempting to characterize and predict aerodynamic loads \citep{Burelle2020}. Moreover, strong non-linear effects emerge under realistic conditions (i.e., for highly-separated, unsteady flows at high Reynolds numbers), contributing to the challenge in the load-estimation process. 

In order to fully account for the effects deriving from such realistic conditions, and to exploit only a sparse set of sensors, existing data-driven methods are continuously improved, as well as novel approaches are proposed. Among other techniques, complex networks represent a powerful and versatile tool for time-series analysis~\citep{zou2019complex} that have been recently employed to study fluid flows \citep{iacobello2020review}, including vortical flows \citep{taira2016network, meena2021identifying}, turbulent-combustor dynamics \citep{krishnan2021Suppression, kobayashi2019early, murugesan2015combustion}, as well as mixing in wall-bounded turbulence \citep{iacobello2019lagrangian, perrone2020wall}. In this context, transition networks -- thanks to their connection with Markov models~\citep{zou2019complex} -- have been successfully employed for time series reconstruction \citep{shirazi2009mapping, campanharo2011duality, mccullough2017regenerating, fernex2020cluster}, as well as for reduced-order modeling \citep{kaiser2014cluster, li2021cluster, foroozan2021unsupervised} and control~\citep{nair2019cluster}. Specifically, Fernex \textit{et al}.~\cite{fernex2020cluster} have recently shown that cluster-based transition networks can be used as an effective data-driven tool to model complex nonlinear dynamical systems (including turbulence) without any prior knowledge.

Despite recent progress, cluster-based transition networks are still predominantly used to \textit{reconstruct} data resulting from accurate, numerically-obtained data. In the reconstruction process, the newly-generated time series are only expected to be globally similar (i.e., sharing similar statistical features) to the reference time series, which is explicitly included in the training dataset \citep{fernex2020cluster, li2021cluster}. In the present study, instead, we apply transition networks to perform signal \textit{estimation} in real-time and with experimentally-obtained data. Here we generate new signals that, based on sparse (sensor) input, estimate the instantaneous aerodynamic state of an aerodynamic body with good local accuracy.

Hereby, three main challenges arise from real-world (experimental) implementations: (i) limited amount of training data (e.g., range of boundary conditions), (ii) sparse data (limited amount of sensors), and (iii) realistic (noisy) data. To tackle these issues, we present an algorithm that -- relying on cluster-based transition networks -- is able to perform signal estimation in highly-separated experimental flows where sparse sensors are available (\figurename \ref{fig:intro2}). In particular, the network-based strategy can exploit sparse datasets as well as the system's dynamics in the recent past for signal estimation, thus mitigating the need to collect large datasets typical of data-driven approaches.

The methodological steps required to build transition networks are reported in section \ref{sec:methods}, where the new network-based strategy allowing for signal estimation is described (section \ref{subsec:methB}). Our strategy is tested on a simple yet challenging, experimental test case of an accelerated elliptical plate (section\,\ref{sec:expmethods}), captured by only two differential pressure sensors. The results of the load estimates are presented and discussed in section\,\ref{sec:results}. Conclusions and future outlook are eventually drawn in section\,\ref{sec:conclusions}.

\section{Transition networks with real-time input}\label{sec:methods}

This section describes the signal estimation process, which exploits the features of a transition network built on an experimental training dataset, and a testing dataset of (input) sparse pressure measurements enforcing a constraint to the estimation process of unknown load signals. As shown in \figurename \ref{fig:intro2}, the overall method is characterized by four main steps: (i) the collection of a training dataset (\figurename \ref{fig:intro2}b); (ii) the definition of a phase space from training data and the construction of transition networks (\figurename \ref{fig:intro2}c); (iii) the measurement of (real-time) input data (\figurename \ref{fig:intro2}d); and (iv) the estimation of the load signal (\figurename \ref{fig:intro2}e). We note that, while the procedural steps leading to the construction of transition networks (see \ref{subsec:training_phasespace}-\ref{subsec:build_transNet}) are mainly based on standard practices in the literature~\cite{kaiser2014cluster, fernex2020cluster, li2021cluster}, the novel strategy adopted here to estimate the load signal is provided in section \ref{subsec:methB}.

\begin{figure*}
	\centering
	\includegraphics[width=0.99\textwidth]{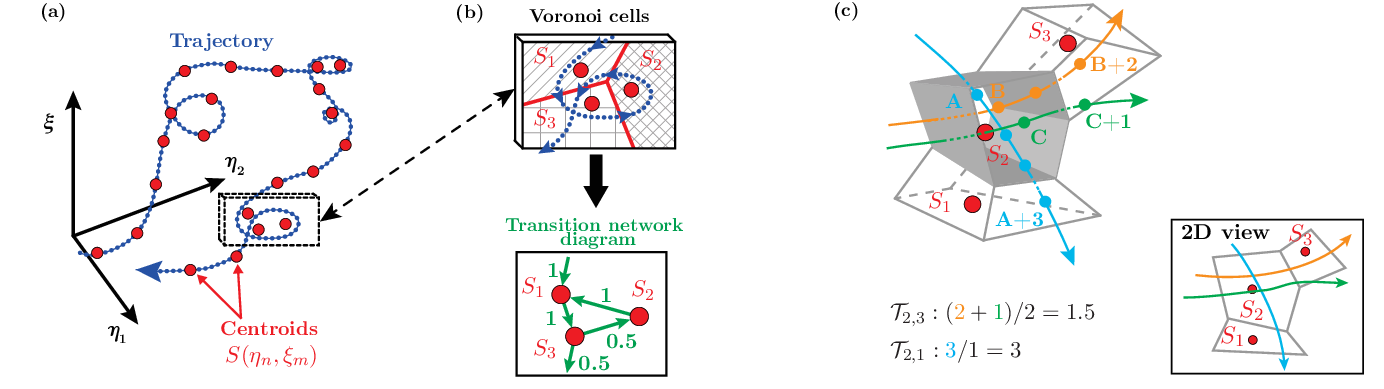}
	\caption{(a) An example of 3D phase space with a trajectory (blue dotted line) and clusters centroids (red dots). (b) A zoomed-in inset in the trajectory, displaying three different Voronoi cells associated to three centroids. The transition network diagram refers to the inset, where green arrows indicate the direction of transition while numbers refer to transition probability values. (c) Schematic for the computation of transition times. The Voronoi cell of $\boldsymbol{S}_2$ is highlighted through gray faces, while the remaining two cells for nodes $\boldsymbol{S}_1$ and $\boldsymbol{S}_3$ are highlight via gray edges.} \label{fig:method_defTransNet}
\end{figure*}

\subsection{Training dataset and phase space clustering}\label{subsec:training_phasespace}

A training dataset consists of $N$ synchronized time series from sparse sensors (here pressure probes) and $M$ signals corresponding to the variables that have to be estimated (here aerodynamic load time series). In general, the training dataset can comprise multiple collections of synchronized time series (see \figurename \ref{fig:intro2}b), where each collection belongs to a different configuration parameter value, $\alpha_i$. The configuration parameter can be, e.g., the Reynolds number, a boundary condition, or a geometrical configuration. In this work, $\alpha$ represents different angles of attack (see section \ref{sec:expmethods}).

In this study, we consider $N=2$ pressure signals and $M=1$ loads (as described in section\,\ref{sec:expmethods}), so that a 3D phase space can be obtained. In a phase space, each variable of the training dataset corresponds to a direction, $\eta_n$ or $\xi_m$ with $n=1,\dots,N$ and $m=1,\dots,M$. \figurename~\ref{fig:method_defTransNet}(a) shows a 3D phase space with directions $\eta_1$, $\eta_2$ and $\xi$, and an exemplifying trajectory depicted as a blue dotted arrow. The rationale behind the phase-space construction is to provide a geometrical representation of a multivariate time series, where each set of values $\lbrace\eta_1(t), \eta_2(t),\xi(t)\rbrace$ at a given time, $t$, indicates a unique dynamical state of the (flow) system through a unique point in the phase space. By mapping signal data at different times into the phase space, an oriented trajectory can then be formed whose direction is in increasing time.

Trajectory points in the phase space are grouped by means of the $k$-means algorithm~\citep{fernex2020cluster, li2021cluster, lloyd1982least,arthur2006k}. Phase-space clustering is usually performed to gain a simplification of the trajectories in the phase space, thus providing a reduced-order representation of the system \cite{li2021cluster, kaiser2014cluster, fernex2020cluster}. Specifically, the $k$-means algorithm partitions the phase space into Voronoi cells represented by their cell centroids, $\boldsymbol{S}^\alpha$, whose entries are the centroid's coordinates in the phase space. The superscript $\bullet^\alpha$ here indicates that the clustering is applied to the trajectory corresponding to the configuration $\alpha$ of the training dataset. As a result, a cluster-based representation of the training dataset is obtained, where each centroid in the phase space represents a coarse-grained aerodynamic state. Following this idea, in this work each centroid represents a specific set of pressure and load values resulting from a specific (unsteady) condition, which in turn is dependent on the angle of attack and its time history).

In the example of \figurename~\ref{fig:method_defTransNet}(a), cluster centroids are depicted as red dots, capturing the essential features of the blue-dotted trajectory. The inset in \figurename~\ref{fig:method_defTransNet}(b) presents three Voronoi cells associated with three centroids, where red straight lines indicate the cell edges. Note that, although the $k$-means performs an unsupervised clustering, it requires an \textit{a priori} definition of the number of clusters (i.e., centroids), $N_{cl}$, which should be large enough to capture the essential geometrical features of the phase-space trajectories. After $N_{cl}$ is fixed, the $k$-means algorithm is applied to each trajectory corresponding to each configuration $\alpha$.

\subsection{Building transition networks}\label{subsec:build_transNet}

Transition networks are constructed from clustered trajectories, where cluster centroids are assigned to network nodes. Accordingly, a univocal correspondence exists between Voronoi cells, their centroids, and network nodes, all indicated through the symbol $\boldsymbol{S}^\alpha$. Network links are weighed on the probability of (temporal) transition between two nodes, thus capturing the temporal dynamics of the complex system. In particular, the probability of transition from node $S_i^\alpha$ to node $S_j^\alpha$ is given by

\begin{equation}
    \mathcal{P}_{i,j}^\alpha=\frac{N(S_i^\alpha,S_j^\alpha)}{N_{all}(S_i^\alpha)}, \quad i,j=1,\dots,N_{cl},
    \label{eq:trans_prob}
\end{equation}

\noindent where $N(S_i^\alpha,S_j^\alpha)$ is the number of times that trajectories directly transit from node $S_i^\alpha$ to node $S_j^\alpha$, while $N_{all}(S_i^\alpha)$ is the total number of times that trajectories exit from node $S_i^\alpha$. In general, $\mathcal{P}_{i,j}$ is not symmetric, i.e., $\mathcal{P}_{i,j}\neq\mathcal{P}_{j,i}$ $\forall i\neq j$, so that network links can be illustrated by means of arrows indicating the direction of transition \citep{newman2018networks}. For example, the blue trajectory in the inset of \figurename~\ref{fig:method_defTransNet}(b) uniquely transits from $S_1$ to $S_3$, so that $\mathcal{P}_{1,3}=1$, as reported in the transition network diagram where links are depicted as green arrows. Moreover, $\sum_{j=1}^{N_{cl}}{\mathcal{P}_{i,j}}=1$ (by definition of probability), and $\mathcal{P}_{i,i}=0$ (by definition of direct transition between nodes\,\citep{li2021cluster}) for any $i=1,\dots,N_{cl}$.

\begin{figure*}
	\centering
	\includegraphics[width=0.97\textwidth]{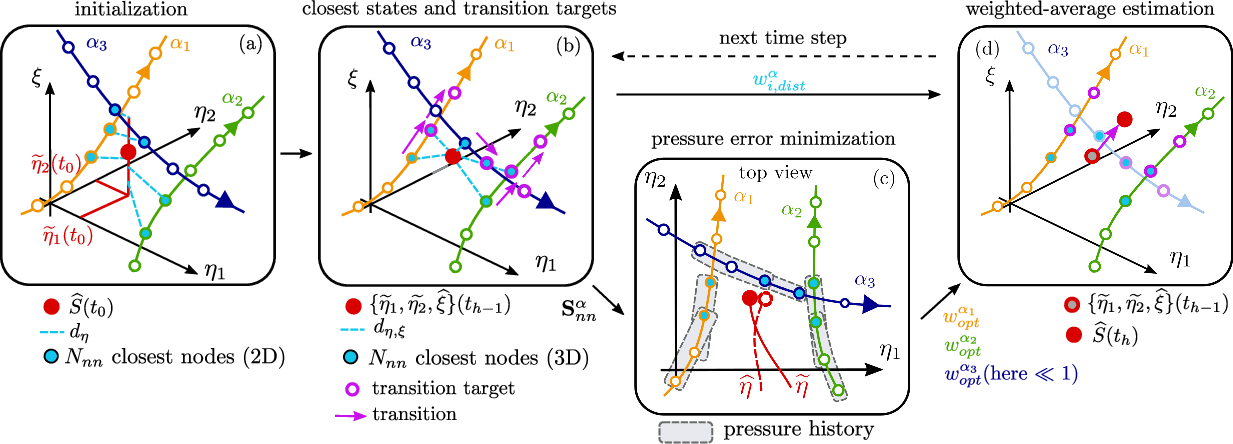}
	\caption{Schematic of the WAB method for a 3D phase space, including three trained trajectories depicted as orange, green, and blue lines. (a) Initial estimate of the load by weighted averaging using the 2D distance $d_\eta$. (b) Identification of transition targets. (c) Weight optimization (optional) based on the recent pressure history. (d) New estimation using transition targets (highlighted in magenta) for the weighted average. Here $w_i^\alpha$ is calculated via eq. (\ref{eq:weights_opt}) using the weights $w_{dist,i}^\alpha$ and  $w^\alpha_{opt}$, obtained in panels (b) and (c), respectively. As in this example the history of $\alpha_3$ is significantly different from the measured data $\widehat{\eta}$, the $\alpha$-dependent weight is small ($w^{\alpha_3}_{opt}\ll 1$). Thus, the impact of the corresponding transition network ($\alpha_3$, transparent in panel (d)) on the estimated state is negligible. }\label{fig:method_B}
\end{figure*}

To fully characterize the transition properties of the network, transition times, $\mathcal{T}_{i,j}^\alpha$, are also defined as the average amount of time needed for the transition from a node $S_i^\alpha$ to a node $S_j^\alpha$ \citep{li2021cluster}. \figurename\,\ref{fig:method_defTransNet}(c) shows a 3D sketch to illustrate the computation of transition times for a given reference cell identified by node $S_2$, where three trajectories (or three intervals of the same trajectory) are illustrated as colored dotted arrows. For node $S_2$, the transition times are $\mathcal{T}_{2,3}=1.5$ and $\mathcal{T}_{2,1}=3$ since the trajectories take, on average, $1.5$ and $3$ time steps to transit from node $S_2$ to $S_3$ and $S_1$, respectively. As per matrix $\boldsymbol{\mathcal{P}}$, the transition times matrix, $\boldsymbol{\mathcal{T}}$, is also generally asymmetric ($\mathcal{T}^\alpha_{i,j}\neq\mathcal{T}^\alpha_{j,i}$).

\subsection{Weighted-average-based (WAB) transition networks}\label{subsec:methB}

The features of the transition-probability matrix, $\boldsymbol{\mathcal{P}}$, and the transition-time matrix, $\boldsymbol{\mathcal{T}}$, can then be used to generate a new set of signals. Owing to the versatility in the transition network construction, in this work we present a strategy to perform signal \textit{estimation} referred to as the weighted-average-based (WAB) transition network. The methodology proposed here performs a weighted average among different states in the phase space to create a trajectory of newly-generated nodes, as well as an optimization procedure that minimizes the difference between the estimated and input (measured) pressure. In particular, here we assume that input values from $N$ time series are known during the signal generation. For example, input values can originate from a set of $N$ sparse pressure sensors collecting data in real time, thereby supporting the time-series estimation (\figurename \ref{fig:intro2}d). 
\\The time series from the testing dataset are hereafter indicated via $\widetilde{\bullet}$ notation (i.e., $\lbrace\widetilde{\eta_1}, \widetilde{\eta_2}, \widetilde{\xi}\rbrace$), while the newly-generated signals are hereafter indicated via $\widehat{\bullet}$ notation (i.e., $\lbrace\widehat{\eta}_1, \widehat{\eta}_2, \widehat{\xi}\rbrace$). We also note that an estimated time vector, $\widehat{t}$, can also be defined since, in general, $\boldsymbol{\mathcal{T}}$ entries do not exactly correspond to the time step $\Delta t$ (a consequence of the clustering operation).

The WAB methodology is described here by highlighting its key features and then sketched in \figurename~\ref{fig:method_B}, while procedural details (due to their elaborated nature) are extensively reported in Appendix \ref{app:wab_details}:

\begin{enumerate}

    \item The first step of WAB is its initialization (\figurename \ref{fig:method_B}a). A load estimation is obtained at the first time, $t_0$, by employing a nearest-neighbor approach, because a transition approach requires at least two times. $N_{nn}$ nearest nodes (see cyan filled circles in \figurename \ref{fig:method_B}a) are identified for each trajectory (corresponding to each $\alpha$) with respect to the measured input pressure at $t_0$. Hence, the load value at $t_0$ is evaluated as the distance-based weighted average of the load values coming from each of the $N_{nn}$ nodes in each trajectory;

    \item The transition probabilities of the networks are then used to continue estimating the load signal at a generic $t_h>t_0$ (\figurename \ref{fig:method_B}b). In particular, $N_{nn}$ nearest nodes, with respect to the pressure input and the previously estimated load, $\lbrace \widetilde{\eta}_1, \widetilde{\eta}_2, \widehat{\xi}\rbrace\left( t_{h-1}\right)$, are first identified for each trajectory. By so doing, a set of weights, $w^{\alpha}_{i,dist}$, can be defined that are inversely proportional to the distance between the $N_{nn}$ closest nodes and $\lbrace \widetilde{\eta}_1, \widetilde{\eta}_2, \widehat{\xi}\rbrace\left( t_{h-1}\right)$ (see cyan dashed lines in \figurename \ref{fig:method_B}b). The transition matrix is then exploited to identify the transition target nodes of each of the $N_{nn}$ (source) nodes, following the criterion of maximum transition probability (see magenta circles and arrows in \figurename \ref{fig:method_B}b);
     
    \item To not only account for the present system state, but also for the recent past of the pressure input, we define a second set of $\alpha$-specific weights, $w^{\alpha}_{opt}$ (see \figurename \ref{fig:method_B}c). The weights, $w^{\alpha}_{opt}$, are generated through an optimization process that minimizes the error between the estimated pressure $\widehat{\eta}$ and the input (measured) pressure $\widetilde{\eta}$ over the recent time period $\Delta t$. As such, the configurations, $\alpha$, with a similar pressure history as the input data, will have a higher impact on the load estimate; and

\begin{figure*}
	\centering
	\includegraphics[width=0.9\textwidth]{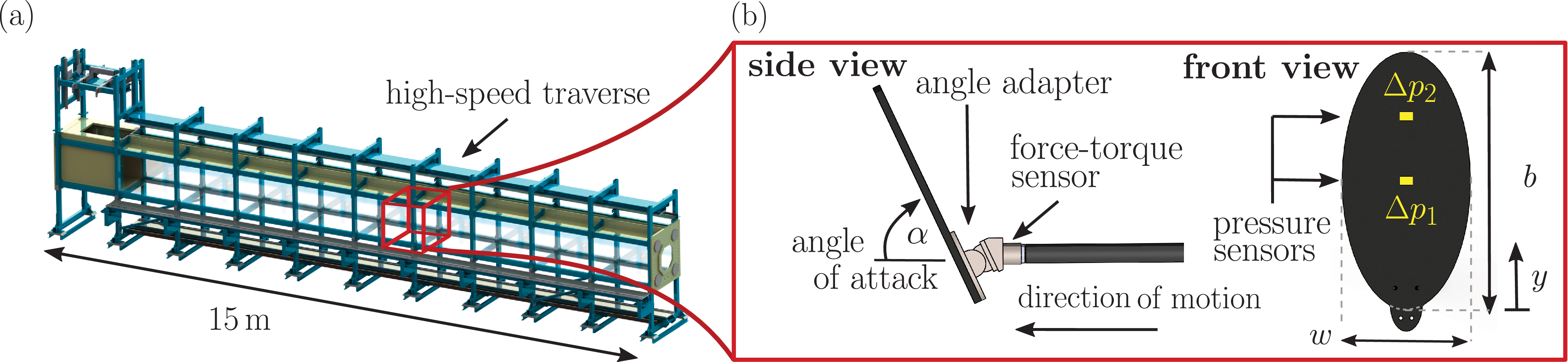}\\\includegraphics[width=0.99\textwidth]{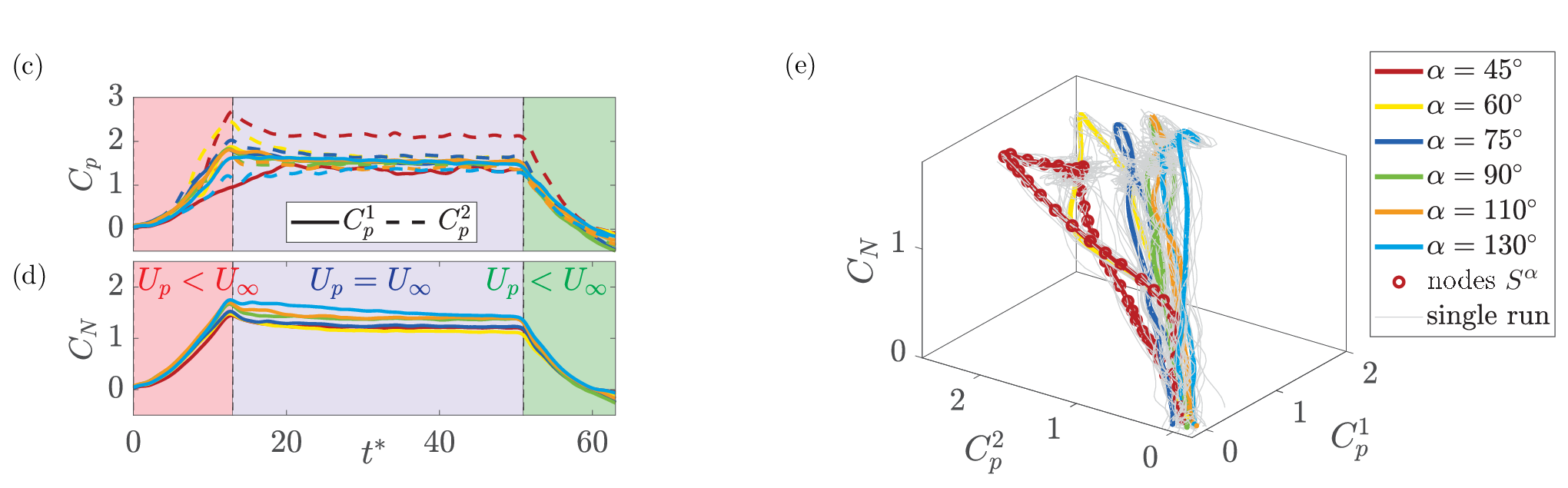}
	\caption{(a) Towing-tank facility; (b) front and side view of the elliptical plate model with force-torque sensor, an angle adapter to determine $\alpha$ and two differential pressure sensors. Normalized plots of (c) the differential pressures, and (d) the plate-normal force as a function of the normalized time $t^*=U_\infty t/D_h$ and for various angles of attack. The acceleration stage (red), steady stage (blue), and the deceleration stage (green) are highlighted. (e) 3D phase space built on $C_N$, $C_p^1$, and $C_p^2$. Phase-averaged data (colored), and single-run data (gray) are both shown. For $\alpha=45^\circ$ (red trajectory), $N_{cl}=60$ nodes, $\boldsymbol{S}^\alpha$ (corresponding to cluster centroids), are also displayed.}
	\label{fig:setup}
\end{figure*}

    \item The load $\widehat{\xi}(t_h)$ is eventually estimated as a weighted average of the loads from the identified target nodes (see red-filled dot in \figurename \ref{fig:method_B}d): 
    	\begin{equation}
		\widehat{\xi}(t_h)=\frac{\sum_\alpha{w^\alpha_{opt}\sum_i{ w_{i,dist}^\alpha  \xi(S^\alpha_i)}}}{\sum_\alpha{w^\alpha_{opt}\sum_i{w^\alpha_{i,dist} }}} \, .		\label{eq:meth_weighAvrg}		
	\end{equation} 
    As such, the weighted average accounts for both the phase-space distances (via $w^{\alpha}_{i,dist}$) and the temporal dynamics  of each trajectory (via $w^{\alpha}_{opt}$).

    To conclude, an estimated time, $\widehat{t}_h$, is also computed by using transition times $\boldsymbol{\mathcal{T}}$ instead of load values in the weighted average.
    
\end{enumerate}


\section{Experimental test case}\label{sec:expmethods}

As a test case for the transition-network frameworks presented in section\,\ref{sec:methods}, realistic experimental flow data were captured in a highly-separated and unsteady flow at a high Reynolds number. In particular, the flow around an accelerating elliptical plate was characterized via pressure and load measurements, and the same experimental set-up was used to obtain both training and testing datasets, as described in section\,\ref{sec:methods}.

\subsection{Test facility and experimental set-up}

The experiments were performed in a fully-enclosed, water-filled (viscosity $\nu$), $15\,$m long towing-tank facility with $1\,\mathrm{m}\times 1\,\mathrm{m}$ cross-section (\figurename \ref{fig:setup}a). The model consisted of an elliptical plate (\figurename \ref{fig:setup}a), with principal axes $b=0.3\,$m and $w=0.15\,$m and a cross-sectional area $A=\pi b w$. The model was connected to the traverse above the towing tank via a horizontal sting with diameter $0.08b$ and length $2b$, and a vertical symmetric profile of thickness $0.08b$. The plate was towed from rest with the plate velocity $U_p$ accelerating at a rate of $0.4$\,$\mathrm{m/s}^{2}$ until hitting its final velocity $U_\infty=1\,$m/s, resulting in a terminal Reynolds number of $Re=U_\infty b/\nu=194\,000$. The plate velocity ($U_p=U_\infty$) was then kept constant over a distance of $\sim 40D_h$ before it was decelerated to rest, where $D_h$ is the hydraulic diameter. The same kinematics were tested for the plate mounted at various angles of attack $\alpha$ (as defined in \figurename \ref{fig:setup}b), in the range of $45^\circ \leq \alpha \leq 130^\circ$. 

Two Omega differential pressure transducers captured the instantaneous differential pressure $\Delta p$ between the two sides of the plate. \figurename \ref{fig:setup}(b) shows the positions of the two pressure transducers at $y/b=0.5$ ($\Delta p_1$) and $y/b=0.75$ ($\Delta p_2$), respectively. The pressure sensors measure a range of $\pm6895\,$Pa, and have a response time of $10^{-3}\,$s, and an accuracy of $\pm 0.25\%$ of the full-scale best fit straight line (FS BFSL) with hysteresis and repeatability of $0.2\%$ FS. In order to measure forces and moments on the plate, an ATI Nano 25 six-axis force-torque sensor was mounted between the plate and the horizontal sting. The transducer has a resolution of $0.125\,$N. Pressures and forces were recorded at a sampling frequency of $1000\,$Hz. 

\figurename \ref{fig:setup}(c-d) presents the temporal evolution of the normalized pressures $C_p^i=2\Delta p_i/(\rho U_\infty^2)$, and the plate-normal load $C_N=2F_N/(\rho A U_\infty^2)$, for the six angles of attack $\alpha = \lbrace 45^\circ, 60^\circ, 75^\circ, 90^\circ, 110^\circ, 130^\circ\rbrace$. Here $i=1,2$ refers to sensor position at $y/b=0.5$ and $y/b=0.75$, respectively. The pressures and loads were phase-averaged over 10 runs, and temporally-filtered with a least-squares estimator \citep{savitzky1964}. \figurename \ref{fig:setup}(e) visualizes the same data of \figurename \ref{fig:setup}(c-d) in 3D phase space, whose directions are $\eta_1=C_p^1$, $\eta_2=C_p^2$ and $\xi=C_N$. Single-run data are also illustrated in \figurename \ref{fig:setup}(e) as gray trajectories, in addition to the phase-averaged data (colored). 
\\For smaller $\alpha$ values, the spacing between different trajectories is notably visible (e.g., the red and yellow trajectories in \figurename \ref{fig:setup}(e)), while similar pressures are observed for $\alpha>75^\circ$, making it difficult in this phase space to distinguish between the trajectories of different $\alpha$. As such, the present dataset is particularly challenging with regard to accurate load estimates using transition networks. Specifically, ambiguous states are likely to appear, namely points in the phase space with similar pressure values but different loads.

\subsection{Transtion-network construction and estimation parameters}

The transition networks were built using a training dataset made up of the pressure data ($\eta_1=C_p^1,\eta_2=C_p^2$) and the plate-normal load ($\xi=C_N$), as well as following the description provided in section\,\ref{sec:methods}. The order of the experimental data was then reduced by clustering the phase-space trajectories for each $\alpha$ into $N_{cl}=300$ centroids, $\boldsymbol{S}^\alpha$. As a representative example, centroids are shown in \figurename~\ref{fig:setup}(e) for the (phase-averaged) trajectory corresponding to the configuration $\alpha=45^\circ$. In general, small values of $N_{cl}$ serve to reduce the computational effort of the method. However, if $N_{cl}$ is too small, the dynamics of a trajectory in the phase space cannot be properly resolved, thus leading to higher estimation errors. In the present study, $N_{cl}=300$ (with 12300 time-series instants, i.e., trajectory points) provides a good balance between estimation accuracy of the load $\widehat{C}_N$ and computational effort. A parametric analysis on the effects of $N_{cl}$ on the results is provided in Appendix\,\ref{app:parametric}.

Once the transition networks are established, a real-time estimate of the plate-normal load $\widehat{\xi}=\widehat{C}_N$ can be obtained by utilizing the pressure sensors' (real-time) input ($\widetilde{\eta}_1=\widetilde{C}_p^1$ and $\widetilde{\eta}_2=\widetilde{C}_p^2$) in combination with the WAB procedure (\ref{subsec:methB}). Specifically, the number of nearest-neighbours $N_{nn}$ (cyan nodes in \figurename~\ref{fig:method_B}a-b) was set equal to $10$. Although $N_{nn}$ is usually set to 3 or 4~\citep{fernex2020cluster}, $N_{nn}=10$ leads to smoother load estimates without substantial changes in the overall results. 
\\Furthermore, pressure-error minimization was performed over a temporal window $\Delta t^*\leq 6$ (i.e., a traveled distance at most equal to 6 times the plate hydraulic diameter). The value of $\Delta t^*$ corresponds to half of the acceleration time $t^*\approx 12$ (see \figurename~\ref{fig:setup}d), and allows us to sufficiently capture the temporal features of the acceleration and deceleration phases. Larger $\Delta t^*$ values do not lead to substantial changes in the results.
\section{Results and Discussion}\label{sec:results}

\begin{figure*}
	\centering
	\includegraphics[width=.99\textwidth]{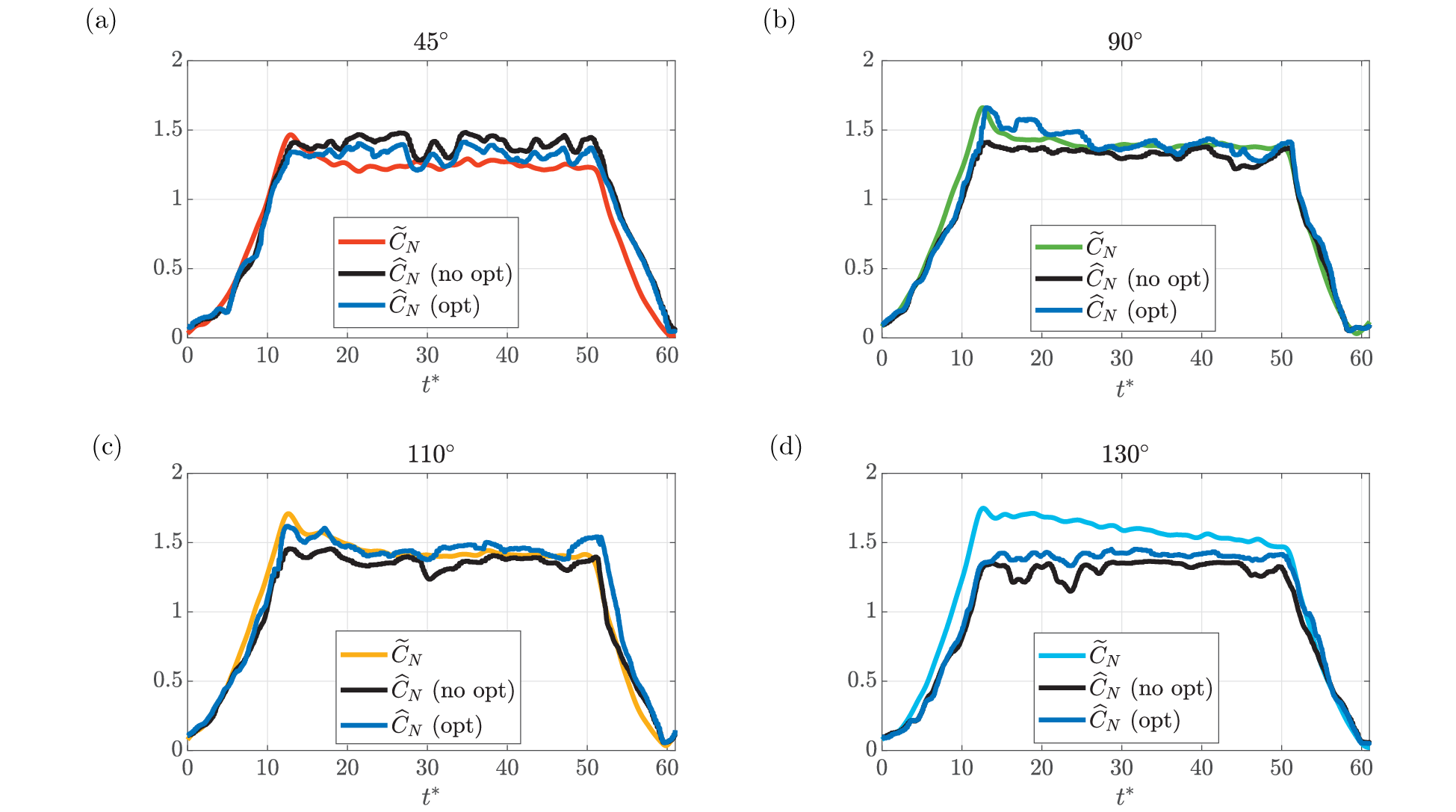}
	\caption{Load estimates for various omitted $\alpha$ values reported in each panel's title. The input pressure data were randomly selected among all single runs of the respective $\alpha$. Estimated loads, $\widehat{C}_N$, with or without pressure-error minimization, are depicted as dark blue and black lines, respectively. As a reference, the measured loads $\widetilde{C}_N$ are shown using the same colors as in \figurename~\ref{fig:setup}(e).}
	\label{fig:result_loads}
\end{figure*}

\begin{figure*}
	\centering
	\includegraphics[width=.9\textwidth]{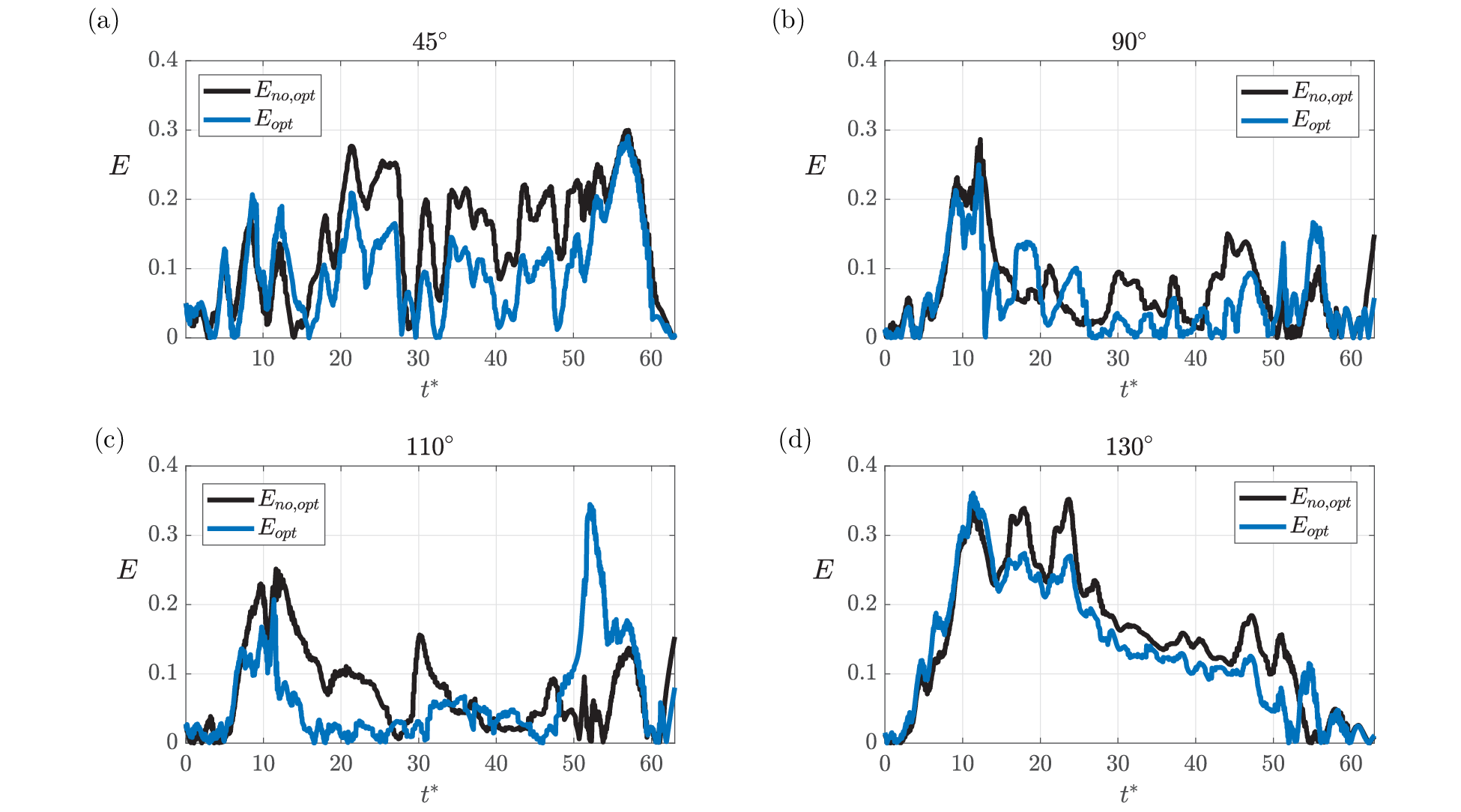}
	\caption{Normalized absolute errors, $E$, for the load estimates shown in \figurename~\ref{fig:result_loads}.}
	\label{fig:result_errors}
\end{figure*}

In this section we present and discuss the results stemming from the application of the WAB approach when the transition networks are used to estimate loads for omitted flow configurations, i.e., time series corresponding to $\alpha$ values that were not available in the training data. To mimic realistic estimation conditions, a randomly-selected single run (gray lines in \figurename~\ref{fig:setup}e) is used as a testing time series, rather than the phase-averaged signals (colored trajectories in \figurename~\ref{fig:setup}e). In this way, we account for single-run noise, as phase-averaged signals are less noisy.

\figurename~\ref{fig:result_loads} shows the resulting load estimates $\widehat{C}_N$ compared to the measured loads $\widetilde{C}_N$, for different omitted $\alpha$ values (reported in each panel's title). To highlight the impact of including the system dynamics in the estimation process, we compare results with and without utilizing the $\alpha$-specific weights $w_{opt}^\alpha$ obtained via the pressure optimization strategy presented in \figurename~\ref{fig:method_B}(c) (dark blue and black lines, respectively). In general, the WAB approach is able to reproduce well the shape of the measured loads $\widetilde{C}_N$. 

To quantify the estimation performance in more detail, \figurename~\ref{fig:result_errors} presents the normalized error, $E = E_{abs}/ \langle\widetilde{C}_N\rangle$, of the estimates $\widehat{C}_N$ with respect to the test data $\widetilde{C}_N$, where $E_{abs}=|\widehat{C}_N-\widetilde{C}_N|$ is the absolute error while $\langle\bullet\rangle$ indicates the time average. The accuracy of the estimate varies depending on the flow stage (i.e., acceleration, steady-state, deceleration; see \figurename~\ref{fig:setup}c-d), the estimation strategy (with or without pressure optimization), as well as on the omitted $\alpha$. The magnitude of $E$ remains relatively small (less than $20\%$) throughout the whole estimation process, even if optimization is not performed (see also \figurename~\ref{fig:appendix_res} for a comprehensive assessment). 
\\Note that, in contrast to previous studies on signal reconstruction~\citep{fernex2020cluster, li2021cluster}, we use the (normalized) absolute error $E$ instead of statistical quantities such as autocorrelation to assess the estimation quality. A discussion on the relevance of statistical similarity in the context of signal estimation is provided in Appendix \ref{app:autocorr}.

Estimating the aerodynamic state in real-world applications such as our accelerated flat plate, imposes several challenges as outlined in section \ref{sec:intro}: (i) a limited amount of training data; (ii) a limited amount of sensors; and (iii) realistic (noisy) data. In the following, we use the present results to discuss how the network-based estimation algorithm presented in section \ref{subsec:methB} tackles those challenges.

\subsection{Limited amount of training data}

Training a data-driven algorithm in an experimental setting comes with significant effort. In fact, experimental campaigns are often time intensive and involve costly facilities. To potentially reduce the required amount of training data needed for accurate load estimates, the WAB approach estimates unknown dynamics by combining information from different configurations, $\alpha$, via a weighted average. A similar approach was proposed by Fernex \textit{et al}.~\cite{fernex2020cluster}, who evaluated the weighted average between two configurations that were identified manually and \textit{a priori}. In contrast, the WAB approach used here takes all available configurations, $\alpha$, of the training dataset into account, since one does not know \textit{a priori} which $\alpha$ value(s) are suitable for the present estimation. 

By omitting a configuration $\alpha$ from the training data, and then estimating the aerodynamic loads of the omitted $\alpha$, we can assess the capabilities of the WAB approach to estimate an unknown signal. As shown in \figurename~\ref{fig:result_loads}, the WAB approach successfully estimates loads from $\alpha$ that were omitted in the training data. The estimates are particularly accurate for $\alpha=90^\circ$ and $\alpha=110^\circ$ (see  \figurename~\ref{fig:result_loads}b,c and \figurename~\ref{fig:result_errors}b,c). However, the performance of the WAB approach deteriorates if $\alpha=130^\circ$ or $\alpha=45^\circ$ are omitted in the training dataset and then estimated (see panels (a) and (d), respectively, in \figurename~\ref{fig:result_loads} and \figurename~\ref{fig:result_errors}). 

Taking $\alpha=130^\circ$ as a representative case, this behavior can be explained by the fact that the trajectory for $\alpha=130^\circ$ is not fully surrounded by other trajectories in the phase space (see \figurename~\ref{fig:setup}e), but is only close to the trajectory for $\alpha=110^\circ$ (orange line in \figurename~\ref{fig:setup}e). From the point of view of WAB, the weighted average to estimate the load for $\alpha=130^\circ$ is performed using load data that are always lower than the expected $\widetilde{C}_N$, so that the average is unavoidably driven by lower $C_N$ values, thus leading to higher estimation errors. In contrast, the trajectories of $\alpha=90^\circ$ and $\alpha=110^\circ$ are surrounded by states of various $\alpha$ (as shown in the phase-space representation of \figurename~\ref{fig:setup}e), so that the WAB strategy can better interpolate to the estimated force.
\\This limitation of the WAB approach is a common feature of interpolation-based techniques, and can be overcome by properly collecting training data (\figurename~\ref{fig:intro2}b) so that all expected peripheral boundary conditions are accounted for. For example, in the present experimental dataset, the training should contain the maximum and minimum $\alpha$ that can be experienced by the system.

\subsection{Limited amount of sensors}\label{sec:limited}

Physically implementing a dense network of pressure sensors on an aerodynamic body requires high design and production costs. However, when the amount of sensors is significantly reduced, ambiguous states will likely occur (i.e., same pressure input values, but different loads; see section \ref{sec:expmethods}). To accurately estimate the aerodynamic loads with sparse data, the present WAB approach concurrently relies on the instantaneous sensor data and the recent history of the system. Namely, the information from the previous state estimate at $t_{h-1}$, and the instantaneous sensor input, are combined to determine the node-specific weights $w_{i,dist}^\alpha$ (\figurename \ref{fig:method_B}b). In addition, the recent pressure history is used to find a set of $\alpha$-specific weights $w^\alpha_{opt}$ that minimize  the error of the pressure estimate within the recent past (window $\Delta t^*$, \figurename \ref{fig:method_B}c). As such, both weights, $w_{i,dist}^\alpha$ and $w^\alpha_{opt}$, contribute to the WAB's performance in systems with sparse sensors.

The positive effects of using $w^\alpha_{opt}$, obtained by the pressure-error minimization, is apparent when comparing the estimated loads with and without optimization in \figurename~\ref{fig:result_loads} and \figurename~\ref{fig:result_errors}. The estimation performance is consistently better when $w^\alpha_{opt}$ is used. This is particularly evident for $\alpha=90^\circ$ and $\alpha=110^\circ$, in which optimized WAB (black lines in \figurename~\ref{fig:result_loads}b-c) is able to capture the initial load bump occurring at the end of the acceleration phase ($t^*\approx 12$). This local increase is particularly challenging to be estimated as a result of the strong unsteadiness and non-linearity in the system, thus highlighting the potential of optimized WAB in performing load estimation effectively.

\subsection{Realistic data}

As shown in section \ref{sec:limited}, including the system dynamics in the estimation process, can lead to better cluster-based modeling and estimation performance. While our WAB approach relies on a pressure-error optimization, alternative approaches were suggested to account for system dynamics during the state estimation. For instance, Nair \textit{et al}.~\cite{nair2019cluster} accounted for the system dynamics by adding the temporal derivative of their (numerical) input data as an additional axis of the phase space, thus obtaining a better aerodynamic state identification. However, under realistic (experimental) conditions, training and input data display noise levels as a result of several (systematic or randomly-appearing) factors affecting the measurements. In particular, the noise level of experimental pressure data obtained in separated, high-Reynolds number flows is typically very high, leading to inaccurate evaluations of temporal gradients. Therefore, in such cases, temporal gradients do not generally represent a suitable choice to be included in the phase space. Accordingly, our WAB approach has been conceived to rely only on absolute values of the sensor input and not on their temporal gradients. 
 
Furthermore, to mitigate the impact of experimental noise on estimation, we used phase-averaged data as a training dataset (see \figurename \ref{fig:setup}c-e). Nevertheless, the estimation capabilities were tested on (randomly-selected) single-run time series, which differ from the phase-averaged signals, and provide additional challenges to the estimation accuracy. In spite of these challenges, the WAB approach still proved to be accurate even in presence of noisy input data.

\section{Conclusions and Outlook}\label{sec:conclusions}

In this study, we extend the application of transition networks from signal reconstruction to load estimation with real-time input. In particular, we generate new signals that were not included in the training dataset, utilizing the input of $N=2$ sparse sensors. A weighted average-based (WAB) network strategy is proposed and tested under realistic conditions of unsteady flows. In particular, the network-based approach is tested on an experimental dataset with pressure and load data from an accelerating elliptical plate at various angles of attack. The WAB strategy exploits the features of transition networks (which comprise the definition of a phase space and a clustering algorithm) and a real-time input of sparse pressure signals. Furthermore, an optimization process that minimizes the difference between estimated and measured (input) pressure is implemented. 

The potential and limitations of the WAB approach are discussed for estimates corresponding to different (omitted) angles of attack. The results indicate that transition networks can estimate configurations that were unknown during the training stage, with global estimation errors below $20\%$, and for some cases even below $10\%$. Moreover, the pressure optimization approach is able to further refine the estimation outcomes by also capturing characteristic local behaviors of the unsteady load signals, e.g., after the acceleration phases. Therefore, our WAB approach proves to be a robust and accurate tool for signal estimation, even in presence of sparse input data and with limited training data.

While the current approach represents a first effort to employ transition networks to estimate aerodynamic loads in unsteady and high-Reynolds number flows, several methodological advancements can be implemented to potentially enhance the capabilities of the WAB approach. In fact, transition networks do not represent a \textit{black box} tool between input and output variables but provide a versatile framework that can be easily modified to account for advanced information on the flow system. 

In this regard, the implementation of the system dynamics through pressure-error minimization represents a paradigmatic example. Additional physical insights can be included in the network model by expanding the size of the phase space, where additional axes could represent other measured variables of the system. In this vein, future efforts aim to incorporate external flow measurements (e.g., velocity or vorticity fields) in the transition network model. On this note, the outcomes from simple models could also be incorporated in the algorithms, either as additional axes of the phase space or as additional input data to constrain the estimation process. 
\\With the aim to account for noise in experimental data, different clustering strategies could also be applied (such as fuzzy algorithms), and the probabilistic nature of transition networks could be further exploited, e.g., implementing Bayesian statistics \citep{kaiser2021AIAA}. Furthermore, different interpolation schemes can be used, where the weighted average can be performed on a limited set of configurations chosen via an optimization routine.

In conclusion, transition networks show great potential for real-time estimation of unknown variables in fluid dynamics problems, even under challenging flow conditions and sparse training datasets. Therefore, we believe the proposed network-based methodology, owing to its versatility, can be a promising tool for the real-time estimation of realistic flows even when limited by sparse data.

\appendix
\section{WAB procedural details}\label{app:wab_details}

The details of the WAB methodology are here reported. We recall that the time series from the testing dataset are indicated via $\widetilde{\bullet}$ notation (i.e., $\lbrace\widetilde{\eta_1}, \widetilde{\eta_2}, \widetilde{\xi}\rbrace$), while the newly-generated signals are hereafter indicated via $\widehat{\bullet}$ notation (i.e., $\lbrace\widehat{\eta}_1, \widehat{\eta}_2, \widehat{\xi}\rbrace$). The WAB approach comprises the following four main steps (see \figurename~\ref{fig:method_B}):

\begin{enumerate}

	\item At the first time, $t_0$, a nearest-neighbor approach is used to estimate $\widehat{\xi}(t_0)$ because a transition approach requires at least two times. First, a set,  $\boldsymbol{S}^\alpha_{nn}$, of $N_{nn}$ nodes is selected for each configuration $\alpha$. Specifically, each set of nodes comprises the closest $N_{nn}$ nodes to the measured pressures of the testing dataset, namely $\boldsymbol{\widetilde{\eta}}(t_0)=\lbrace\widetilde{\eta}_1(t_0), \widetilde{\eta}_2(t_0)\rbrace$. For example, in \figurename~\ref{fig:method_B}(a), $N_{nn}=2$ and the closest nodes to the two configurations $\alpha_1$ and $\alpha_2$ are identified by dashed cyan lines, which highlight the 2D distances $d_\eta=||\boldsymbol{\eta}(\boldsymbol{S}^\alpha_{nn})- \boldsymbol{\widetilde{\eta}}(t_0)||_2$. The load value, $\widehat{\xi}(t_0)$, is then evaluated as the weighted average of the $\xi$ values of each node in $\boldsymbol{S}^\alpha_{nn}$, namely 
				
	\begin{equation}
		\widehat{\xi}(t_0)=\frac{\sum_\alpha{\sum_i{w_i^\alpha \xi(S^\alpha_i)}}}{\sum_\alpha{\sum_i{w_i^\alpha}}},		\label{eq:meth_weighAvrg}		
	\end{equation} 
				
	\noindent where $S^\alpha_i\in\boldsymbol{S}^\alpha_{nn}$, while $w_i^\alpha=1/d_\eta$ is a set of distance-dependent weights. A newly-estimated node $\boldsymbol{\widehat{S}}(t_0)=\lbrace \widetilde{\eta_1}(t_0), \widetilde{\eta_2}(t_0), \widehat{\xi}(t_0) \rbrace$ is then obtained, which is illustrated as a filled red dot in \figurename~\ref{fig:method_B}(a); \label{item:methB_step1}

\begin{figure*}
	\centering
    \includegraphics[width=0.99\textwidth]{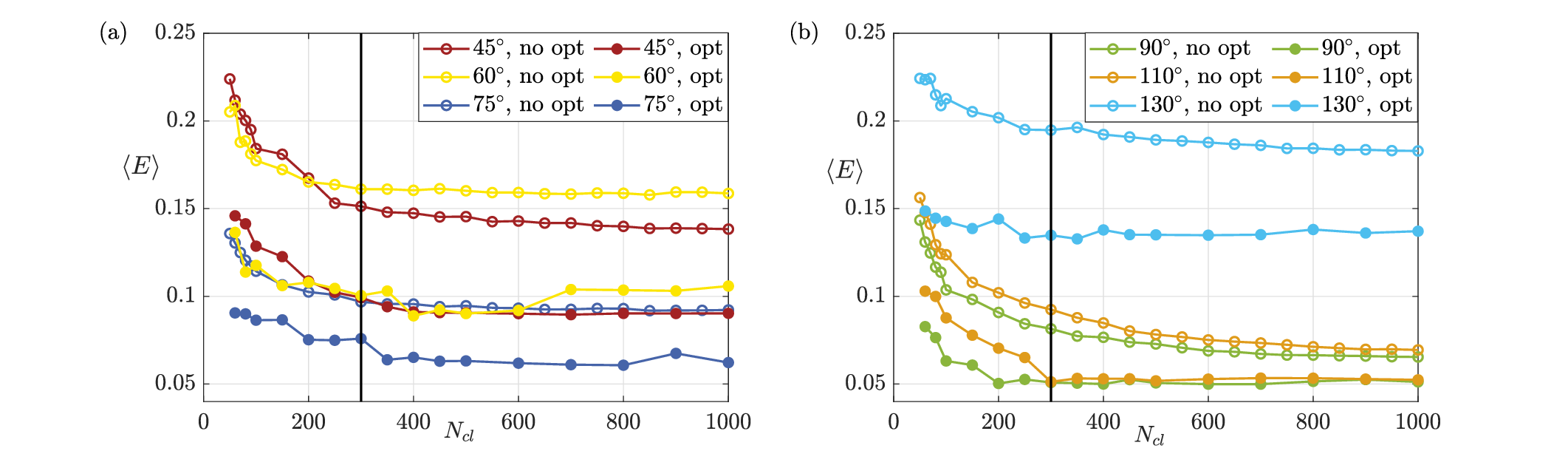}
	\caption{Effect on the number nodes (i.e., cluster centroids) on the estimation performance. The legend indicates the estimated case: (a) $\alpha=45^\circ, 60^\circ, 75^\circ$; (b) $\alpha=90^\circ, 110^\circ, 130^\circ$. The black vertical line corresponds to $N_{cl}=300$ as in section~\ref{sec:results}.}
	\label{fig:appendix_res}
\end{figure*}

	\item The transition probabilities of the networks are then used to continue estimating the load signal. At a generic time $t_h$, the $N_{nn}$ closest nodes, $\boldsymbol{S}_{nn}^\alpha$, are first identified. 
	In particular, the nodes in $\boldsymbol{S}_{nn}^\alpha$ are selected to minimize the Euclidean distance, $d_{\eta,\xi}$, between the point $\lbrace \widetilde{\eta}_1, \widetilde{\eta}_2, \widehat{\xi}\rbrace\left( t_{h-1}\right)$ (see filled red circle in \figurename~\ref{fig:method_B}b) and all the nodes belonging to each trajectory. For any node in $\boldsymbol{S}_{nn}^\alpha$, the transition matrix is exploited to identify the transition target nodes following the criterion of maximum transition probability. Target nodes are highlighted by magenta circles in each trajectory of \figurename~\ref{fig:method_B}(b), while transitions are shown via magenta arrows. 
	
	Similar to the initialization at $t_0$ (\figurename \ref{fig:method_B}a), a set of weights $w^\alpha_{dist,i}=1/d_{\eta,\xi}$ can be defined, implying that closer $\boldsymbol{S}_{nn}^\alpha$ nodes will have a higher impact (i.e., a higher weight) on the estimation of the next state $\widehat{S}(t_{h})$. We note that during initialization a 2D distance ($d_\eta$) was used. For $t_h>t_0$ a load estimate $\widehat{\xi}(t_h)$ exists that can be exploited to calculate a 3D distance ($d_{\eta,\xi}$), as shown by cyan dashed lines in \figurename~\ref{fig:method_B}(b). In contrast to a 2D distance, the 3D distance provides more robustness against ambiguous estimations. These arise when the $\eta_1$ and $\eta_2$ values of the nodes $\boldsymbol{S}_{nn}^\alpha$ are similar, but their $\xi$ values are significantly different;\label{item:methB_step2}

	\item Although the identified nodes $\boldsymbol{S}_{nn}^\alpha$ are close to $\widehat{S}(t_{h-1})$ in the phase space, this does not necessarily imply that $\boldsymbol{S}_{nn}^\alpha$ nodes belong to a trajectory (i.e., an $\alpha$ case) displaying a similar temporal dynamics as the input data $\widetilde{\boldsymbol{\eta}}$. Therefore, to identify the $\alpha$ cases that best capture the dynamics of the input data, we rely on the recent past of our input data $\widetilde{\boldsymbol{\eta}}$.
	In particular, we define a new set of $\alpha$-specific coefficients, $w_{opt}^\alpha$, which are generated through a pressure-error minimization strategy (\figurename~\ref{fig:method_B}c).
	The input data, $\widetilde{\boldsymbol{\eta}}$, in a temporal window $[t_{h-\delta h},t_h]$ are used as reference values and compared with the estimated pressure values $\widehat{\boldsymbol{\eta}}$, computed as
	\begin{equation}
	    \widehat{\boldsymbol{\eta}}=\frac{\sum_\alpha{\sum_i{w_i^\alpha \boldsymbol{\eta}(S^\alpha_i)}}}{\sum_\alpha{\sum_i{w_i^\alpha}}},	
	\end{equation}
	where
		\begin{equation}
	    w_i^\alpha=w_{i,dist}^\alpha\cdot w_{opt}^\alpha. \label{eq:weights_opt}
	\end{equation}

	A minimization problem is then solved which aims to find the set of weights $w_{opt}^\alpha$ that minimizes the maximum absolute difference between input and estimated pressure (dashed and solid red lines in figure \ref{fig:method_B}c) over the chosen temporal window, $[t_{h-\delta h},t_h]$, namely
	 \begin{equation}\label{eq:opt_min}
	    \arg\min_{w_{opt}^\alpha} \left[\max_{t\in [t_{h-\delta h},t_h]} \left[|\widehat{\boldsymbol{\eta}}_{1,2}(t;w_{opt}^\alpha)- \widetilde{\boldsymbol{\eta}}_{1,2}(t)|\right]\right].
	\end{equation}
	
	\noindent We note that $w_{opt}^\alpha$ only depends on $\alpha$, thus providing a measure of the reliability of each trajectory (corresponding to configurations $\alpha$) to fulfill the constrain on estimated pressure coming from input (testing) data. For example, the blue trajectory corresponding to $\alpha_3$ in \figurename~\ref{fig:method_B}(c) is much less reliable than the remaining two trajectories (for $\alpha_1$ and $\alpha_2$) in providing a good estimation for pressure, thus leading to $w_{opt}^{\alpha_3}\ll 1$. If optimization is not performed, $w_{opt}^\alpha=1$ for any $\alpha$ configuration and $w_i^\alpha\equiv w_{i,dist}^\alpha$; and

	\item Finally, a weighted average of the $\xi$ values of the target nodes is computed to estimate the load value $\widehat{\xi}$ at time $t_h$, thus obtaining the point $\boldsymbol{\widehat{S}}(t_h)$ (\figurename~\ref{fig:method_B}d). Equation (\ref{eq:meth_weighAvrg}) is still exploited to get $\boldsymbol{\widehat{S}}(t_h)$, but the weights $w_i^\alpha=w_{i,dist}^\alpha\cdot w_{opt}^\alpha$ (eq. \ref{eq:weights_opt}) are defined as the product of the distance-related weight $w_{i,dist}^\alpha=1/d_{\eta,\xi}$ (cyan dashed lines in \figurename~\ref{fig:method_B}b) and the $\alpha$-specific novel coefficient $w_{opt}^\alpha$ (\figurename~\ref{fig:method_B}c) which accounts for the system dynamics.

	 
	 To conclude the procedure, the estimated time $\widehat{t}_h$ is also computed as the sum of the previous estimated time, $\widehat{t}_{h-1}$, and a weighted-average transition time from eq. (\ref{eq:meth_weighAvrg}) where transition times $\boldsymbol{\mathcal{T}}$ are used instead of $\xi$.

\end{enumerate}

 \begin{figure*}
	\centering
	\includegraphics[width=.45\textwidth]{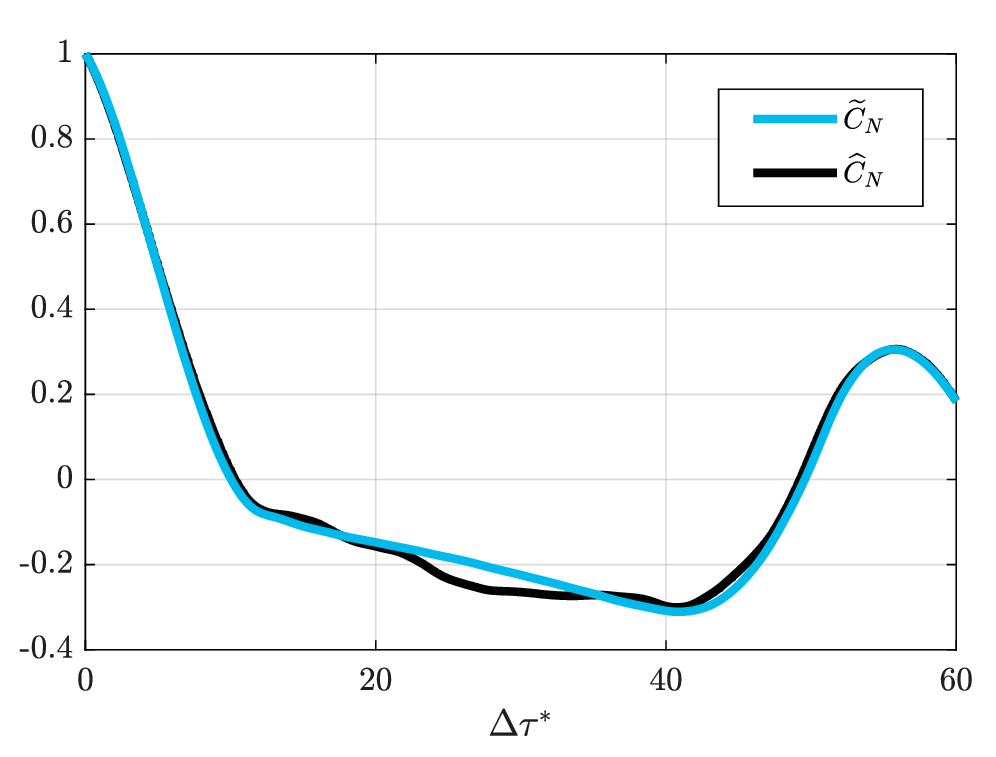}
	\caption{Autocorrelogram for the measured (cyan line) and estimated (black line) loads for omitted case $\alpha=130^\circ$. Estimated load is obtained without optimization. The root-mean-square error between autocorrelations is $0.022$.}
	\label{fig:autocorr_result}
\end{figure*}

\section{Parametric analysis on number of network nodes}\label{app:parametric}

This Appendix describes the effects the $N_{cl}$ parameter on the estimation performances of the WAB transition network strategy. We recall that $N_{cl}$ indicates the number of nodes in the network, which correspond to the centroids of the Voronoi cells obtained from the $k$-means clustering.

\figurename \ref{fig:appendix_res} shows the average performance of WAB when $N_{cl}$ is varied, either with or without pressure-error minimization. Here a global error is computed as $\langle E\rangle=\langle E_{abs}\rangle/\langle\widetilde{C}_N\rangle$. In general, $\langle E\rangle$ increases towards small $N_{cl}$ values because $N_{cl}$ becomes comparable with the number of nodes used to perform the weighted average, $N_{nn}=10$. As discussed in section~\ref{sec:results}, the WAB method performs well when intermediate configurations have to be estimated, which is confirmed in \figurename \ref{fig:appendix_res} for $\alpha=\lbrace 75^\circ, 90^\circ, 110^\circ\rbrace$ (see blue, green and orange lines respectively). In particular, it is evident as the pressure-error minimization (filled-dotted lines) always reduces the overall estimation error for any omitted $\alpha$. Finally we highlight that $\langle E\rangle$ remains quite constant for $N_{cl}>300$, with values below $10\%$ for the intermediate cases and below $20\%$ for external cases (i.e., $\alpha=45^\circ,130^\circ$), thus justifying the choice of $N_{cl}=300$ in section~\ref{sec:results}.

\section{Reconstruction versus estimation}
\label{app:autocorr}

Data driven approaches have often been  used to \textit{reconstruct} data, so that the newly-generated reconstructed time series are expected to share very similar statistical features with respect to the corresponding signal included in the training dataset \citep{fernex2020cluster, li2021cluster}. In contrast, in the present study, we aimed to \textit{estimate} unknown signals, i.e., not included in the training dataset. In general, the newly estimated signals could display very similar statistical features with respect to the expected time-series, while still containing considerable local errors. In other words, although an estimation process could produce a globally (statistically) similar time-series with respect to the expected signal, local errors can be non-negligible. 

In our work, this could be a consequence of the fact that each single run will always differ locally from other runs or from phase-average signals, that are included in the training dataset. A representative example is provided in \figurename~\ref{fig:autocorr_result}: we show as a black line the autocorrelogram of the estimated load $\widehat{C}_N$ (without performing optimization) for the omitted case $\alpha=130^\circ$, as a function of the temporal lag $\Delta \tau^*$. For comparison, the autocorrelogram of the measured (reference) load $\widetilde{C}_N$ is also reported as a cyan line. Autocorrelation is chosen here in analogy with previous studies assessing the capabilities of transition networks~\citep{fernex2020cluster, li2021cluster}. While estimation errors can be locally non-negligible (as illustrated in panel (d) of \figurename~\ref{fig:result_loads} and \figurename~\ref{fig:result_errors}), the difference between the autocorrelogram for the estimated and measured loads is instead very small. 

Therefore, while statistical tools like the autocorrelation could be effectively used to assess \textit{reconstruction} performances, they might not always provide a reliable measure of the \textit{estimation} performances, especially in the context of unsteady load estimation.


%
%

%


\bibliography{biblio_list}

\providecommand{\noopsort}[1]{}\providecommand{\singleletter}[1]{#1}%
\begin{thebibliography}{31}%
\makeatletter
\providecommand \@ifxundefined [1]{%
 \@ifx{#1\undefined}
}%
\providecommand \@ifnum [1]{%
 \ifnum #1\expandafter \@firstoftwo
 \else \expandafter \@secondoftwo
 \fi
}%
\providecommand \@ifx [1]{%
 \ifx #1\expandafter \@firstoftwo
 \else \expandafter \@secondoftwo
 \fi
}%
\providecommand \natexlab [1]{#1}%
\providecommand \enquote  [1]{``#1''}%
\providecommand \bibnamefont  [1]{#1}%
\providecommand \bibfnamefont [1]{#1}%
\providecommand \citenamefont [1]{#1}%
\providecommand \href@noop [0]{\@secondoftwo}%
\providecommand \href [0]{\begingroup \@sanitize@url \@href}%
\providecommand \@href[1]{\@@startlink{#1}\@@href}%
\providecommand \@@href[1]{\endgroup#1\@@endlink}%
\providecommand \@sanitize@url [0]{\catcode `\\12\catcode `\$12\catcode
  `\&12\catcode `\#12\catcode `\^12\catcode `\_12\catcode `\%12\relax}%
\providecommand \@@startlink[1]{}%
\providecommand \@@endlink[0]{}%
\providecommand \url  [0]{\begingroup\@sanitize@url \@url }%
\providecommand \@url [1]{\endgroup\@href {#1}{\urlprefix }}%
\providecommand \urlprefix  [0]{URL }%
\providecommand \Eprint [0]{\href }%
\providecommand \doibase [0]{https://doi.org/}%
\providecommand \selectlanguage [0]{\@gobble}%
\providecommand \bibinfo  [0]{\@secondoftwo}%
\providecommand \bibfield  [0]{\@secondoftwo}%
\providecommand \translation [1]{[#1]}%
\providecommand \BibitemOpen [0]{}%
\providecommand \bibitemStop [0]{}%
\providecommand \bibitemNoStop [0]{.\EOS\space}%
\providecommand \EOS [0]{\spacefactor3000\relax}%
\providecommand \BibitemShut  [1]{\csname bibitem#1\endcsname}%
\let\auto@bib@innerbib\@empty
\bibitem [{\citenamefont {Liao}\ \emph {et~al.}(2003)\citenamefont {Liao},
  \citenamefont {Beal}, \citenamefont {Lauder},\ and\ \citenamefont
  {Triantafyllou}}]{liao2003fish}%
  \BibitemOpen
  \bibfield  {author} {\bibinfo {author} {\bibfnamefont {J.}~\bibnamefont
  {Liao}}, \bibinfo {author} {\bibfnamefont {D.}~\bibnamefont {Beal}}, \bibinfo
  {author} {\bibfnamefont {G.}~\bibnamefont {Lauder}},\ and\ \bibinfo {author}
  {\bibfnamefont {M.}~\bibnamefont {Triantafyllou}},\ }\bibfield  {title}
  {\enquote {\bibinfo {title} {Fish exploiting vortices decrease muscle
  activity},}\ }\href@noop {} {\bibfield  {journal} {\bibinfo  {journal}
  {Science}\ }\textbf {\bibinfo {volume} {302}},\ \bibinfo {pages} {1566--1569}
  (\bibinfo {year} {2003})}\BibitemShut {NoStop}%
\bibitem [{\citenamefont {Portugal}\ \emph {et~al.}(2014)\citenamefont
  {Portugal}, \citenamefont {Hubel}, \citenamefont {Fritz}, \citenamefont
  {Heese}, \citenamefont {Trobe}, \citenamefont {Voelkl}, \citenamefont
  {Hailes}, \citenamefont {Wilson},\ and\ \citenamefont
  {Usherwood}}]{portugal2014upwash}%
  \BibitemOpen
  \bibfield  {author} {\bibinfo {author} {\bibfnamefont {S.}~\bibnamefont
  {Portugal}}, \bibinfo {author} {\bibfnamefont {T.}~\bibnamefont {Hubel}},
  \bibinfo {author} {\bibfnamefont {J.}~\bibnamefont {Fritz}}, \bibinfo
  {author} {\bibfnamefont {S.}~\bibnamefont {Heese}}, \bibinfo {author}
  {\bibfnamefont {D.}~\bibnamefont {Trobe}}, \bibinfo {author} {\bibfnamefont
  {B.}~\bibnamefont {Voelkl}}, \bibinfo {author} {\bibfnamefont
  {S.}~\bibnamefont {Hailes}}, \bibinfo {author} {\bibfnamefont {A.~M.}\
  \bibnamefont {Wilson}},\ and\ \bibinfo {author} {\bibfnamefont {J.~R.}\
  \bibnamefont {Usherwood}},\ }\bibfield  {title} {\enquote {\bibinfo {title}
  {Upwash exploitation and downwash avoidance by flap phasing in ibis formation
  flight},}\ }\href@noop {} {\bibfield  {journal} {\bibinfo  {journal}
  {Nature}\ }\textbf {\bibinfo {volume} {505}},\ \bibinfo {pages} {399--402}
  (\bibinfo {year} {2014})}\BibitemShut {NoStop}%
\bibitem [{\citenamefont {Zbikowski}(2004)}]{zbikowski2004}%
  \BibitemOpen
  \bibfield  {author} {\bibinfo {author} {\bibfnamefont {R.}~\bibnamefont
  {Zbikowski}},\ }\bibfield  {title} {\enquote {\bibinfo {title} {Sensor-rich
  feedback control: a new paradigm for flight control inspired by insect
  agility},}\ }\href@noop {} {\bibfield  {journal} {\bibinfo  {journal} {IEEE
  Instru. Meas. Mag.}\ }\textbf {\bibinfo {volume} {7}},\ \bibinfo {pages}
  {19--26} (\bibinfo {year} {2004})}\BibitemShut {NoStop}%
\bibitem [{\citenamefont {Sterbing-D'Angelo}\ \emph {et~al.}(2011)\citenamefont
  {Sterbing-D'Angelo}, \citenamefont {Chadha}, \citenamefont {Chiu},
  \citenamefont {Falk}, \citenamefont {Xian}, \citenamefont {Barcelo},
  \citenamefont {Zook},\ and\ \citenamefont {Moss}}]{sterbing2011}%
  \BibitemOpen
  \bibfield  {author} {\bibinfo {author} {\bibfnamefont {S.}~\bibnamefont
  {Sterbing-D'Angelo}}, \bibinfo {author} {\bibfnamefont {M.}~\bibnamefont
  {Chadha}}, \bibinfo {author} {\bibfnamefont {C.}~\bibnamefont {Chiu}},
  \bibinfo {author} {\bibfnamefont {B.}~\bibnamefont {Falk}}, \bibinfo {author}
  {\bibfnamefont {W.}~\bibnamefont {Xian}}, \bibinfo {author} {\bibfnamefont
  {J.}~\bibnamefont {Barcelo}}, \bibinfo {author} {\bibfnamefont {J.~M.}\
  \bibnamefont {Zook}},\ and\ \bibinfo {author} {\bibfnamefont {C.~F.}\
  \bibnamefont {Moss}},\ }\bibfield  {title} {\enquote {\bibinfo {title} {Bat
  wing sensors support flight control},}\ }\href@noop {} {\bibfield  {journal}
  {\bibinfo  {journal} {PNAS USA}\ }\textbf {\bibinfo {volume} {108}},\
  \bibinfo {pages} {11291--11296} (\bibinfo {year} {2011})}\BibitemShut
  {NoStop}%
\bibitem [{\citenamefont {Saini}, \citenamefont {Narsipur},\ and\ \citenamefont
  {Gopalarathnam}(2021)}]{saini2021leading}%
  \BibitemOpen
  \bibfield  {author} {\bibinfo {author} {\bibfnamefont {A.}~\bibnamefont
  {Saini}}, \bibinfo {author} {\bibfnamefont {S.}~\bibnamefont {Narsipur}},\
  and\ \bibinfo {author} {\bibfnamefont {A.}~\bibnamefont {Gopalarathnam}},\
  }\bibfield  {title} {\enquote {\bibinfo {title} {Leading-edge flow sensing
  for detection of vortex shedding from airfoils in unsteady flows},}\
  }\href@noop {} {\bibfield  {journal} {\bibinfo  {journal} {Phys. Fluids}\
  }\textbf {\bibinfo {volume} {33}},\ \bibinfo {pages} {087105} (\bibinfo
  {year} {2021})}\BibitemShut {NoStop}%
\bibitem [{\citenamefont {Le~Provost}, \citenamefont {He},\ and\ \citenamefont
  {Williams}(2018)}]{provost2018}%
  \BibitemOpen
  \bibfield  {author} {\bibinfo {author} {\bibfnamefont {M.}~\bibnamefont
  {Le~Provost}}, \bibinfo {author} {\bibfnamefont {X.}~\bibnamefont {He}},\
  and\ \bibinfo {author} {\bibfnamefont {D.~R.}\ \bibnamefont {Williams}},\
  }\bibfield  {title} {\enquote {\bibinfo {title} {Real-time roll and pitching
  moment identification with distributed surface pressure sensors on a ucas
  wing},}\ }in\ \href@noop {} {\emph {\bibinfo {booktitle} {2018 AIAA Aerosp.
  Sci. Meet.}}}\ (\bibinfo {year} {2018})\ p.\ \bibinfo {pages}
  {0326}\BibitemShut {NoStop}%
\bibitem [{\citenamefont {Burelle}\ \emph {et~al.}(2020)\citenamefont
  {Burelle}, \citenamefont {Yang}, \citenamefont {Kaiser},\ and\ \citenamefont
  {Rival}}]{Burelle2020}%
  \BibitemOpen
  \bibfield  {author} {\bibinfo {author} {\bibfnamefont {L.}~\bibnamefont
  {Burelle}}, \bibinfo {author} {\bibfnamefont {W.}~\bibnamefont {Yang}},
  \bibinfo {author} {\bibfnamefont {F.}~\bibnamefont {Kaiser}},\ and\ \bibinfo
  {author} {\bibfnamefont {D.~E.}\ \bibnamefont {Rival}},\ }\bibfield  {title}
  {\enquote {\bibinfo {title} {Exploring the signature of distributed pressure
  measurements on non-slender delta wings during axial and vertical gusts},}\
  }\href@noop {} {\bibfield  {journal} {\bibinfo  {journal} {Phys. Fluids}\
  }\textbf {\bibinfo {volume} {32}} (\bibinfo {year} {2020})}\BibitemShut
  {NoStop}%
\bibitem [{\citenamefont {Wood}\ \emph {et~al.}(2019)\citenamefont {Wood},
  \citenamefont {Araujo-Estrada}, \citenamefont {Richardson},\ and\
  \citenamefont {Windsor}}]{wood2019}%
  \BibitemOpen
  \bibfield  {author} {\bibinfo {author} {\bibfnamefont {K.~T.}\ \bibnamefont
  {Wood}}, \bibinfo {author} {\bibfnamefont {S.}~\bibnamefont
  {Araujo-Estrada}}, \bibinfo {author} {\bibfnamefont {T.}~\bibnamefont
  {Richardson}},\ and\ \bibinfo {author} {\bibfnamefont {S.}~\bibnamefont
  {Windsor}},\ }\bibfield  {title} {\enquote {\bibinfo {title} {Distributed
  pressure sensing--based flight control for small fixed-wing unmanned aerial
  systems},}\ }\href@noop {} {\bibfield  {journal} {\bibinfo  {journal} {J.
  Aircr.}\ }\textbf {\bibinfo {volume} {56}},\ \bibinfo {pages} {1951--1960}
  (\bibinfo {year} {2019})}\BibitemShut {NoStop}%
\bibitem [{\citenamefont {Hou}, \citenamefont {Darakananda},\ and\
  \citenamefont {Eldredge}(2019)}]{hou2019}%
  \BibitemOpen
  \bibfield  {author} {\bibinfo {author} {\bibfnamefont {W.}~\bibnamefont
  {Hou}}, \bibinfo {author} {\bibfnamefont {D.}~\bibnamefont {Darakananda}},\
  and\ \bibinfo {author} {\bibfnamefont {J.}~\bibnamefont {Eldredge}},\
  }\bibfield  {title} {\enquote {\bibinfo {title} {Machine-learning-based
  detection of aerodynamic disturbances using surface pressure measurements},}\
  }\href@noop {} {\bibfield  {journal} {\bibinfo  {journal} {AIAA J.}\ }\textbf
  {\bibinfo {volume} {57}},\ \bibinfo {pages} {5079--5093} (\bibinfo {year}
  {2019})}\BibitemShut {NoStop}%
\bibitem [{\citenamefont {Zou}\ \emph {et~al.}(2019)\citenamefont {Zou},
  \citenamefont {Donner}, \citenamefont {Marwan}, \citenamefont {Donges},\ and\
  \citenamefont {Kurths}}]{zou2019complex}%
  \BibitemOpen
  \bibfield  {author} {\bibinfo {author} {\bibfnamefont {Y.}~\bibnamefont
  {Zou}}, \bibinfo {author} {\bibfnamefont {R.}~\bibnamefont {Donner}},
  \bibinfo {author} {\bibfnamefont {N.}~\bibnamefont {Marwan}}, \bibinfo
  {author} {\bibfnamefont {J.}~\bibnamefont {Donges}},\ and\ \bibinfo {author}
  {\bibfnamefont {J.}~\bibnamefont {Kurths}},\ }\bibfield  {title} {\enquote
  {\bibinfo {title} {Complex network approaches to nonlinear time series
  analysis},}\ }\href@noop {} {\bibfield  {journal} {\bibinfo  {journal} {Phys.
  Rep.}\ }\textbf {\bibinfo {volume} {787}},\ \bibinfo {pages} {1--97}
  (\bibinfo {year} {2019})}\BibitemShut {NoStop}%
\bibitem [{\citenamefont {Iacobello}, \citenamefont {Ridolfi},\ and\
  \citenamefont {Scarsoglio}(2020)}]{iacobello2020review}%
  \BibitemOpen
  \bibfield  {author} {\bibinfo {author} {\bibfnamefont {G.}~\bibnamefont
  {Iacobello}}, \bibinfo {author} {\bibfnamefont {L.}~\bibnamefont {Ridolfi}},\
  and\ \bibinfo {author} {\bibfnamefont {S.}~\bibnamefont {Scarsoglio}},\
  }\bibfield  {title} {\enquote {\bibinfo {title} {A review on turbulent and
  vortical flow analyses via complex networks},}\ }\href@noop {} {\bibfield
  {journal} {\bibinfo  {journal} {Phys. A}\ ,\ \bibinfo {pages} {125476}}
  (\bibinfo {year} {2020})}\BibitemShut {NoStop}%
\bibitem [{\citenamefont {Taira}, \citenamefont {Nair},\ and\ \citenamefont
  {Brunton}(2016)}]{taira2016network}%
  \BibitemOpen
  \bibfield  {author} {\bibinfo {author} {\bibfnamefont {K.}~\bibnamefont
  {Taira}}, \bibinfo {author} {\bibfnamefont {A.}~\bibnamefont {Nair}},\ and\
  \bibinfo {author} {\bibfnamefont {S.}~\bibnamefont {Brunton}},\ }\bibfield
  {title} {\enquote {\bibinfo {title} {Network structure of two-dimensional
  decaying isotropic turbulence},}\ }\href@noop {} {\bibfield  {journal}
  {\bibinfo  {journal} {J. Fluid Mech.}\ }\textbf {\bibinfo {volume} {795}}
  (\bibinfo {year} {2016})}\BibitemShut {NoStop}%
\bibitem [{\citenamefont {Gopalakrishnan~Meena}\ and\ \citenamefont
  {Taira}(2021)}]{meena2021identifying}%
  \BibitemOpen
  \bibfield  {author} {\bibinfo {author} {\bibfnamefont {M.}~\bibnamefont
  {Gopalakrishnan~Meena}}\ and\ \bibinfo {author} {\bibfnamefont
  {K.}~\bibnamefont {Taira}},\ }\bibfield  {title} {\enquote {\bibinfo {title}
  {Identifying vortical network connectors for turbulent flow modification},}\
  }\href@noop {} {\bibfield  {journal} {\bibinfo  {journal} {J. Fluid Mech.}\
  }\textbf {\bibinfo {volume} {915}},\ \bibinfo {pages} {A10} (\bibinfo {year}
  {2021})}\BibitemShut {NoStop}%
\bibitem [{\citenamefont {Krishnan}\ \emph {et~al.}(2021)\citenamefont
  {Krishnan}, \citenamefont {Sujith}, \citenamefont {Marwan},\ and\
  \citenamefont {Kurths}}]{krishnan2021Suppression}%
  \BibitemOpen
  \bibfield  {author} {\bibinfo {author} {\bibfnamefont {A.}~\bibnamefont
  {Krishnan}}, \bibinfo {author} {\bibfnamefont {R.}~\bibnamefont {Sujith}},
  \bibinfo {author} {\bibfnamefont {N.}~\bibnamefont {Marwan}},\ and\ \bibinfo
  {author} {\bibfnamefont {J.}~\bibnamefont {Kurths}},\ }\bibfield  {title}
  {\enquote {\bibinfo {title} {Suppression of thermoacoustic instability by
  targeting the hubs of the turbulent networks in a bluff body stabilized
  combustor},}\ }\href@noop {} {\bibfield  {journal} {\bibinfo  {journal} {J.
  Fluid Mech.}\ }\textbf {\bibinfo {volume} {916}},\ \bibinfo {pages} {A20}
  (\bibinfo {year} {2021})}\BibitemShut {NoStop}%
\bibitem [{\citenamefont {Kobayashi}\ \emph {et~al.}(2019)\citenamefont
  {Kobayashi}, \citenamefont {Murayama}, \citenamefont {Hachijo},\ and\
  \citenamefont {Gotoda}}]{kobayashi2019early}%
  \BibitemOpen
  \bibfield  {author} {\bibinfo {author} {\bibfnamefont {T.}~\bibnamefont
  {Kobayashi}}, \bibinfo {author} {\bibfnamefont {S.}~\bibnamefont {Murayama}},
  \bibinfo {author} {\bibfnamefont {T.}~\bibnamefont {Hachijo}},\ and\ \bibinfo
  {author} {\bibfnamefont {H.}~\bibnamefont {Gotoda}},\ }\bibfield  {title}
  {\enquote {\bibinfo {title} {Early detection of thermoacoustic combustion
  instability using a methodology combining complex networks and machine
  learning},}\ }\href@noop {} {\bibfield  {journal} {\bibinfo  {journal} {Phys.
  Rev. Appl.}\ }\textbf {\bibinfo {volume} {11}},\ \bibinfo {pages} {064034}
  (\bibinfo {year} {2019})}\BibitemShut {NoStop}%
\bibitem [{\citenamefont {Murugesan}\ and\ \citenamefont
  {Sujith}(2015)}]{murugesan2015combustion}%
  \BibitemOpen
  \bibfield  {author} {\bibinfo {author} {\bibfnamefont {M.}~\bibnamefont
  {Murugesan}}\ and\ \bibinfo {author} {\bibfnamefont {R.}~\bibnamefont
  {Sujith}},\ }\bibfield  {title} {\enquote {\bibinfo {title} {Combustion noise
  is scale-free: transition from scale-free to order at the onset of
  thermoacoustic instability},}\ }\href@noop {} {\bibfield  {journal} {\bibinfo
   {journal} {J. Fluid Mech.}\ }\textbf {\bibinfo {volume} {772}},\ \bibinfo
  {pages} {225--245} (\bibinfo {year} {2015})}\BibitemShut {NoStop}%
\bibitem [{\citenamefont {Iacobello}\ \emph {et~al.}(2019)\citenamefont
  {Iacobello}, \citenamefont {Scarsoglio}, \citenamefont {Kuerten},\ and\
  \citenamefont {Ridolfi}}]{iacobello2019lagrangian}%
  \BibitemOpen
  \bibfield  {author} {\bibinfo {author} {\bibfnamefont {G.}~\bibnamefont
  {Iacobello}}, \bibinfo {author} {\bibfnamefont {S.}~\bibnamefont
  {Scarsoglio}}, \bibinfo {author} {\bibfnamefont {J.}~\bibnamefont
  {Kuerten}},\ and\ \bibinfo {author} {\bibfnamefont {L.}~\bibnamefont
  {Ridolfi}},\ }\bibfield  {title} {\enquote {\bibinfo {title} {Lagrangian
  network analysis of turbulent mixing},}\ }\href@noop {} {\bibfield  {journal}
  {\bibinfo  {journal} {J. Fluid Mech.}\ }\textbf {\bibinfo {volume} {865}},\
  \bibinfo {pages} {546--562} (\bibinfo {year} {2019})}\BibitemShut {NoStop}%
\bibitem [{\citenamefont {Perrone}\ \emph {et~al.}(2020)\citenamefont
  {Perrone}, \citenamefont {Kuerten}, \citenamefont {Ridolfi},\ and\
  \citenamefont {Scarsoglio}}]{perrone2020wall}%
  \BibitemOpen
  \bibfield  {author} {\bibinfo {author} {\bibfnamefont {D.}~\bibnamefont
  {Perrone}}, \bibinfo {author} {\bibfnamefont {J.}~\bibnamefont {Kuerten}},
  \bibinfo {author} {\bibfnamefont {L.}~\bibnamefont {Ridolfi}},\ and\ \bibinfo
  {author} {\bibfnamefont {S.}~\bibnamefont {Scarsoglio}},\ }\bibfield  {title}
  {\enquote {\bibinfo {title} {Wall-induced anisotropy effects on turbulent
  mixing in channel flow: A network-based analysis},}\ }\href@noop {}
  {\bibfield  {journal} {\bibinfo  {journal} {Phys. Rev. E}\ }\textbf {\bibinfo
  {volume} {102}},\ \bibinfo {pages} {043109} (\bibinfo {year}
  {2020})}\BibitemShut {NoStop}%
\bibitem [{\citenamefont {Shirazi}\ \emph {et~al.}(2009)\citenamefont
  {Shirazi}, \citenamefont {Jafari}, \citenamefont {Davoudi}, \citenamefont
  {Peinke}, \citenamefont {Tabar},\ and\ \citenamefont
  {Sahimi}}]{shirazi2009mapping}%
  \BibitemOpen
  \bibfield  {author} {\bibinfo {author} {\bibfnamefont {A.}~\bibnamefont
  {Shirazi}}, \bibinfo {author} {\bibfnamefont {G.}~\bibnamefont {Jafari}},
  \bibinfo {author} {\bibfnamefont {J.}~\bibnamefont {Davoudi}}, \bibinfo
  {author} {\bibfnamefont {J.}~\bibnamefont {Peinke}}, \bibinfo {author}
  {\bibfnamefont {M.}~\bibnamefont {Tabar}},\ and\ \bibinfo {author}
  {\bibfnamefont {M.}~\bibnamefont {Sahimi}},\ }\bibfield  {title} {\enquote
  {\bibinfo {title} {Mapping stochastic processes onto complex networks},}\
  }\href@noop {} {\bibfield  {journal} {\bibinfo  {journal} {J. Stat. Mech.}\
  }\textbf {\bibinfo {volume} {2009}},\ \bibinfo {pages} {P07046} (\bibinfo
  {year} {2009})}\BibitemShut {NoStop}%
\bibitem [{\citenamefont {Campanharo}\ \emph {et~al.}(2011)\citenamefont
  {Campanharo}, \citenamefont {Sirer}, \citenamefont {Malmgren}, \citenamefont
  {Ramos},\ and\ \citenamefont {Amaral}}]{campanharo2011duality}%
  \BibitemOpen
  \bibfield  {author} {\bibinfo {author} {\bibfnamefont {A.}~\bibnamefont
  {Campanharo}}, \bibinfo {author} {\bibfnamefont {M.~I.}\ \bibnamefont
  {Sirer}}, \bibinfo {author} {\bibfnamefont {R.~D.}\ \bibnamefont {Malmgren}},
  \bibinfo {author} {\bibfnamefont {F.~M.}\ \bibnamefont {Ramos}},\ and\
  \bibinfo {author} {\bibfnamefont {L.}~\bibnamefont {Amaral}},\ }\bibfield
  {title} {\enquote {\bibinfo {title} {Duality between time series and
  networks},}\ }\href@noop {} {\bibfield  {journal} {\bibinfo  {journal} {PloS
  one}\ }\textbf {\bibinfo {volume} {6}},\ \bibinfo {pages} {e23378} (\bibinfo
  {year} {2011})}\BibitemShut {NoStop}%
\bibitem [{\citenamefont {McCullough}\ \emph {et~al.}(2017)\citenamefont
  {McCullough}, \citenamefont {Sakellariou}, \citenamefont {Stemler},\ and\
  \citenamefont {Small}}]{mccullough2017regenerating}%
  \BibitemOpen
  \bibfield  {author} {\bibinfo {author} {\bibfnamefont {M.}~\bibnamefont
  {McCullough}}, \bibinfo {author} {\bibfnamefont {K.}~\bibnamefont
  {Sakellariou}}, \bibinfo {author} {\bibfnamefont {T.}~\bibnamefont
  {Stemler}},\ and\ \bibinfo {author} {\bibfnamefont {M.}~\bibnamefont
  {Small}},\ }\bibfield  {title} {\enquote {\bibinfo {title} {Regenerating time
  series from ordinal networks},}\ }\href@noop {} {\bibfield  {journal}
  {\bibinfo  {journal} {Chaos}\ }\textbf {\bibinfo {volume} {27}},\ \bibinfo
  {pages} {035814} (\bibinfo {year} {2017})}\BibitemShut {NoStop}%
\bibitem [{\citenamefont {Fernex}, \citenamefont {Noack},\ and\ \citenamefont
  {Semaan}(2021)}]{fernex2020cluster}%
  \BibitemOpen
  \bibfield  {author} {\bibinfo {author} {\bibfnamefont {D.}~\bibnamefont
  {Fernex}}, \bibinfo {author} {\bibfnamefont {B.~R.}\ \bibnamefont {Noack}},\
  and\ \bibinfo {author} {\bibfnamefont {R.}~\bibnamefont {Semaan}},\
  }\bibfield  {title} {\enquote {\bibinfo {title} {Cluster-based network
  modeling—from snapshots to complex dynamical systems},}\ }\href@noop {}
  {\bibfield  {journal} {\bibinfo  {journal} {Sci. Adv.}\ }\textbf {\bibinfo
  {volume} {7}},\ \bibinfo {pages} {eabf5006} (\bibinfo {year}
  {2021})}\BibitemShut {NoStop}%
\bibitem [{\citenamefont {Kaiser}\ \emph {et~al.}(2014)\citenamefont {Kaiser},
  \citenamefont {Noack}, \citenamefont {Cordier}, \citenamefont {Spohn},
  \citenamefont {Segond}, \citenamefont {Abel}, \citenamefont {Daviller},
  \citenamefont {{\"O}sth}, \citenamefont {Krajnovi{\'c}},\ and\ \citenamefont
  {Niven}}]{kaiser2014cluster}%
  \BibitemOpen
  \bibfield  {author} {\bibinfo {author} {\bibfnamefont {E.}~\bibnamefont
  {Kaiser}}, \bibinfo {author} {\bibfnamefont {B.}~\bibnamefont {Noack}},
  \bibinfo {author} {\bibfnamefont {L.}~\bibnamefont {Cordier}}, \bibinfo
  {author} {\bibfnamefont {A.}~\bibnamefont {Spohn}}, \bibinfo {author}
  {\bibfnamefont {M.}~\bibnamefont {Segond}}, \bibinfo {author} {\bibfnamefont
  {M.}~\bibnamefont {Abel}}, \bibinfo {author} {\bibfnamefont {G.}~\bibnamefont
  {Daviller}}, \bibinfo {author} {\bibfnamefont {J.}~\bibnamefont {{\"O}sth}},
  \bibinfo {author} {\bibfnamefont {S.}~\bibnamefont {Krajnovi{\'c}}},\ and\
  \bibinfo {author} {\bibfnamefont {R.}~\bibnamefont {Niven}},\ }\bibfield
  {title} {\enquote {\bibinfo {title} {Cluster-based reduced-order modelling of
  a mixing layer},}\ }\href@noop {} {\bibfield  {journal} {\bibinfo  {journal}
  {J. Fluid Mech.}\ }\textbf {\bibinfo {volume} {754}},\ \bibinfo {pages}
  {365--414} (\bibinfo {year} {2014})}\BibitemShut {NoStop}%
\bibitem [{\citenamefont {Li}\ \emph {et~al.}(2021)\citenamefont {Li},
  \citenamefont {Fernex}, \citenamefont {Semaan}, \citenamefont {Tan},
  \citenamefont {Morzy{\'n}ski},\ and\ \citenamefont {Noack}}]{li2021cluster}%
  \BibitemOpen
  \bibfield  {author} {\bibinfo {author} {\bibfnamefont {H.}~\bibnamefont
  {Li}}, \bibinfo {author} {\bibfnamefont {D.}~\bibnamefont {Fernex}}, \bibinfo
  {author} {\bibfnamefont {R.}~\bibnamefont {Semaan}}, \bibinfo {author}
  {\bibfnamefont {J.}~\bibnamefont {Tan}}, \bibinfo {author} {\bibfnamefont
  {M.}~\bibnamefont {Morzy{\'n}ski}},\ and\ \bibinfo {author} {\bibfnamefont
  {B.~R.}\ \bibnamefont {Noack}},\ }\bibfield  {title} {\enquote {\bibinfo
  {title} {Cluster-based network model},}\ }\href@noop {} {\bibfield  {journal}
  {\bibinfo  {journal} {J. Fluid Mech.}\ }\textbf {\bibinfo {volume} {906}}
  (\bibinfo {year} {2021})}\BibitemShut {NoStop}%
\bibitem [{\citenamefont {Foroozan}\ \emph {et~al.}(2021)\citenamefont
  {Foroozan}, \citenamefont {Guerrero}, \citenamefont {Ianiro},\ and\
  \citenamefont {Discetti}}]{foroozan2021unsupervised}%
  \BibitemOpen
  \bibfield  {author} {\bibinfo {author} {\bibfnamefont {F.}~\bibnamefont
  {Foroozan}}, \bibinfo {author} {\bibfnamefont {V.}~\bibnamefont {Guerrero}},
  \bibinfo {author} {\bibfnamefont {A.}~\bibnamefont {Ianiro}},\ and\ \bibinfo
  {author} {\bibfnamefont {S.}~\bibnamefont {Discetti}},\ }\bibfield  {title}
  {\enquote {\bibinfo {title} {Unsupervised modelling of a transitional
  boundary layer},}\ }\href@noop {} {\bibfield  {journal} {\bibinfo  {journal}
  {J. Fluid Mech.}\ }\textbf {\bibinfo {volume} {929}},\ \bibinfo {pages} {A3}
  (\bibinfo {year} {2021})}\BibitemShut {NoStop}%
\bibitem [{\citenamefont {Nair}\ \emph {et~al.}(2019)\citenamefont {Nair},
  \citenamefont {Yeh}, \citenamefont {Kaiser}, \citenamefont {Noack},
  \citenamefont {Brunton},\ and\ \citenamefont {Taira}}]{nair2019cluster}%
  \BibitemOpen
  \bibfield  {author} {\bibinfo {author} {\bibfnamefont {A.}~\bibnamefont
  {Nair}}, \bibinfo {author} {\bibfnamefont {C.}~\bibnamefont {Yeh}}, \bibinfo
  {author} {\bibfnamefont {E.}~\bibnamefont {Kaiser}}, \bibinfo {author}
  {\bibfnamefont {B.}~\bibnamefont {Noack}}, \bibinfo {author} {\bibfnamefont
  {S.}~\bibnamefont {Brunton}},\ and\ \bibinfo {author} {\bibfnamefont
  {K.}~\bibnamefont {Taira}},\ }\bibfield  {title} {\enquote {\bibinfo {title}
  {Cluster-based feedback control of turbulent post-stall separated flows},}\
  }\href@noop {} {\bibfield  {journal} {\bibinfo  {journal} {J. Fluid Mech.}\
  }\textbf {\bibinfo {volume} {875}},\ \bibinfo {pages} {345--375} (\bibinfo
  {year} {2019})}\BibitemShut {NoStop}%
\bibitem [{\citenamefont {Lloyd}(1982)}]{lloyd1982least}%
  \BibitemOpen
  \bibfield  {author} {\bibinfo {author} {\bibfnamefont {S.}~\bibnamefont
  {Lloyd}},\ }\bibfield  {title} {\enquote {\bibinfo {title} {Least squares
  quantization in pcm},}\ }\href@noop {} {\bibfield  {journal} {\bibinfo
  {journal} {IEEE T. Inform. Theory}\ }\textbf {\bibinfo {volume} {28}},\
  \bibinfo {pages} {129--137} (\bibinfo {year} {1982})}\BibitemShut {NoStop}%
\bibitem [{\citenamefont {Arthur}\ and\ \citenamefont
  {Vassilvitskii}(2006)}]{arthur2006k}%
  \BibitemOpen
  \bibfield  {author} {\bibinfo {author} {\bibfnamefont {D.}~\bibnamefont
  {Arthur}}\ and\ \bibinfo {author} {\bibfnamefont {S.}~\bibnamefont
  {Vassilvitskii}},\ }\href@noop {} {\enquote {\bibinfo {title} {k-means++: The
  advantages of careful seeding},}\ }\bibinfo {type} {Tech. Rep.}\ (\bibinfo
  {institution} {Stanford},\ \bibinfo {year} {2006})\BibitemShut {NoStop}%
\bibitem [{\citenamefont {Newman}(2018)}]{newman2018networks}%
  \BibitemOpen
  \bibfield  {author} {\bibinfo {author} {\bibfnamefont {M.}~\bibnamefont
  {Newman}},\ }\href@noop {} {\emph {\bibinfo {title} {Networks}}}\ (\bibinfo
  {publisher} {Oxford University Press},\ \bibinfo {year} {2018})\BibitemShut
  {NoStop}%
\bibitem [{\citenamefont {Savitzky}\ and\ \citenamefont
  {Golay}(1964)}]{savitzky1964}%
  \BibitemOpen
  \bibfield  {author} {\bibinfo {author} {\bibfnamefont {A.}~\bibnamefont
  {Savitzky}}\ and\ \bibinfo {author} {\bibfnamefont {M.~J.~E.}\ \bibnamefont
  {Golay}},\ }\bibfield  {title} {\enquote {\bibinfo {title} {Smoothing and
  differentiation of data by simplified least squares procedures},}\
  }\href@noop {} {\bibfield  {journal} {\bibinfo  {journal} {Anal. Chem.}\
  }\textbf {\bibinfo {volume} {36}},\ \bibinfo {pages} {1627--1639} (\bibinfo
  {year} {1964})}\BibitemShut {NoStop}%
\bibitem [{\citenamefont {Kaiser}, \citenamefont {Iacobello},\ and\
  \citenamefont {Rival}(2022)}]{kaiser2021AIAA}%
  \BibitemOpen
  \bibfield  {author} {\bibinfo {author} {\bibfnamefont {F.}~\bibnamefont
  {Kaiser}}, \bibinfo {author} {\bibfnamefont {G.}~\bibnamefont {Iacobello}},\
  and\ \bibinfo {author} {\bibfnamefont {D.~E.}\ \bibnamefont {Rival}},\
  }\bibfield  {title} {\enquote {\bibinfo {title} {Aerodynamic state estimation
  from sparse sensor data by pairing bayesian statistics with transition
  networks},}\ }in\ \href@noop {} {\emph {\bibinfo {booktitle} {Accepted at
  2022 AIAA Aerosp. Sci. Meet.}}}\ (\bibinfo {year} {2022})\BibitemShut
  {NoStop}%
\end{thebibliography}%

\end{document}